\documentclass[10pt,journal,compsoc]{IEEEtran}
\usepackage{amsmath,amssymb,amsfonts}
\usepackage{algorithmic}
\usepackage{graphicx}
\usepackage{textcomp}
\usepackage{xcolor}
\usepackage{array}

\usepackage[normalem]{ulem}
\usepackage{booktabs} % For formal tables
\usepackage{enumitem}

\makeatletter
\newcommand{\srcsize}{\@setfontsize{\srcsize}{5pt}{5pt}}
\makeatother

\usepackage[skip=2pt]{caption}
\usepackage{array}
\usepackage{textcomp}

\usepackage{listings}
\usepackage{xcolor}

\definecolor{darkgreen}{rgb}{0,0.5,0}
\definecolor{tinygray}{rgb}{0.9,0.9,0.9}

\newcommand\blue[1]{\textcolor{black}{#1}}
\newcommand\red[1]{\textcolor{black}{#1}}

\makeatletter
\g@addto@macro\normalsize{%
  \setlength\abovedisplayskip{1pt}
  \setlength\belowdisplayskip{1pt}
  \setlength\abovedisplayshortskip{1pt}
  \setlength\belowdisplayshortskip{1pt}
}
\makeatother

\ifCLASSOPTIONcompsoc
  \usepackage[nocompress]{cite}
\else
  \usepackage{cite}
\fi
\ifCLASSINFOpdf
\else
\fi
\begin{document}
\title{Accelerating Binarized Neural Networks via Bit-Tensor-Cores in Turing GPUs}

\author{Ang~Li,~\IEEEmembership{Member,~IEEE,} and~Simon~Su,~\IEEEmembership{Member,~IEEE}% <-this % stops a space
\IEEEcompsocitemizethanks{\IEEEcompsocthanksitem A. Li is a computer scientist from the High-performance Computing group of Pacific Northwest National Laboratory (PNNL), Richland, WA, USA.\protect\\
E-mail: ang.li@pnnl.gov, see http://www.angliphd.com
\IEEEcompsocthanksitem S. Su is a computer scientist from the DoD Supercomputing Resource Center of U.S. Army Research Laboratory (ARL), Aberdeen Proving Ground, MD, USA.}% <-this % stops an unwanted space
}

\IEEEtitleabstractindextext{%
\begin{abstract}
Despite foreseeing tremendous speedups over conventional deep neural networks, the performance advantage of binarized neural networks (BNNs) has merely been showcased on general-purpose processors such as CPUs and GPUs. In fact, due to being unable to leverage bit-level-parallelism with a word-based architecture, GPUs have been criticized for extremely low utilization (1\%) when executing BNNs. Consequently, the latest tensorcores in NVIDIA Turing GPUs start to experimentally support bit computation. In this work, we look into this brand new bit computation capability and characterize its unique features. We show that the stride of memory access can significantly affect performance delivery and a data-format co-design is highly desired to support the tensorcores for achieving superior performance than existing software solutions without tensorcores. We realize the tensorcore-accelerated BNN design, particularly the major functions for fully-connect and convolution layers --- bit matrix multiplication and bit convolution. Evaluations on two NVIDIA Turing GPUs show that, with ResNet-18, our BTC-BNN design can process ImageNet at a rate of 5.6K images per second, 77\% faster than state-of-the-art. Our BNN approach is released on https://github.com/pnnl/TCBNN.
\end{abstract}}

\maketitle

\IEEEdisplaynontitleabstractindextext

% \fi
\IEEEpeerreviewmaketitle



\IEEEraisesectionheading{\section{Introduction}\label{sec:introduction}}

\IEEEPARstart{B}{inarized}-neural-network (BNN) \cite{courbariaux2016binarized, hubara2016binarized, rastegari2016xnor} is an alternative type of deep-neural-networks (DNNs). Compared to general DNNs, such as multi-layer-perceptrons (MLPs) and convolution-neural-networks (CNNs), the major difference of BNN is that it uses a single bit to represent each entry of the input and weight matrices. BNN evolved from DNN through binarized-weight-network (BWN) \cite{courbariaux2015binaryconnect}. It was firstly observed that if the weight matrix can be binarized to $+1$ and $-1$, the floating-point (FP) multiplications can be degraded to addition (i.e., \texttt{mul} $+1$) and subtraction (i.e., \texttt{mul} $-1$). Later, it was further observed that if the input matrix can be binarized as well, then even the floating-point additions and subtractions in BWN can be degraded to logical operations (i.e., \texttt{xnor} for bit dot-product and \texttt{popc} for bit accumulation) \cite{courbariaux2016binarized, hubara2016binarized, rastegari2016xnor}.

BNNs bring several advantages over the full-precision DNNs: (a) \emph{Reduced and simplified computation}. Through binarization, each segment of 32 FP fused-multiply-add (FMA) operations can be aggregated into an \texttt{xnor} operation and a \texttt{popc} operation, leading to theoretically 16$\times$ speedups; (b) \emph{Reduced data movement and storage}. Through binarization, the whole memory hierarchy and network, including registers, caches, scratchpad, DRAM, NoC, etc. can accommodate 32$\times$ in both bandwidth and capacity; (c) \emph{Reduced cost} which comprises energy reduction from simplified hardware design and smaller chip area; (d) \emph{Resilience}. It has been reported that compared with differentiable DNNs, the discrete BNNs exhibit superior stability and robustness against adversarial attacks \cite{galloway2017attacking, khalil2018combinatorial}.

On the flip side of the coin, binarization reduces the model's capacity and discretizes the parameter space, leading to certain accuracy loss. With the tremendous effort from the machine learning (ML) community \cite{hubara2016binarized, tang2017train, darabi2018bnn, lahoud2019self}, accuracy of BNNs have been dramatically improved. The top-1 training accuracy of BNN-based AlexNet and ResNet-18 on ImageNet dataset has achieved 46.1$\%$ \cite{darabi2018bnn} and 56.4$\%$ \cite{liu2018bi}  (54.3$\%$ and 61$\%$ with boosting \cite{zhu2018binary}), with respect to 56.6$\%$ and 69.3$\%$ for full-precision DNN \cite{bethge2019binarydensenet}. A latest BNN work even reported a top-1 accuracy of 70.7$\%$ \cite{bethge2020meliusnet}. 

%So far, the best BNN training accuracy for ImageNet has been achieved by MeliusNet , which impressively demonstrates a top-1 accuracy of 70.7$\%$.

%many HPC \cite{denby1990neural, likhomanenko2015lhcb, baldi2016parameterized, say2018planning, korneev2018constrained, ma2018binary}

%cloud applications \cite{covington2016deep, he2017neural, hauswald2015djinn, kannan2019grandslam}

%baldi2016parameterized

Although BNN is not likely to substitute DNNs because of reduced model capacity, for many HPC \cite{say2018planning, korneev2018constrained, ma2018binary} and cloud applications \cite{covington2016deep, he2017neural}, when certain accuracy levels can be achieved, alternative factors such as latency, energy, hardware cost, resilience, etc. become more prominent. This is especially the case for actual deployment \cite{chen2020gpu}.

%The latest BNN training accuracy is achieved by MeliusNet \cite{} ("MeliusNet: Can Binary Neural Networks Achieve MobileNet-level Accuracy?"), with ImageNet Top-1 accuracy 70.7$\%$, with respect to MobileNet-1.0 with $70.6\%$.

%, with respect to 70.6$\%$ using full-precision MobileNet. 

Despite featuring various advantages, the expected performance gain of BNN has rarely been demonstrated on general purpose processors such as GPUs. This is mainly because: (i) the fundamental design mismatch between bit-based algorithms and word-based architecture; (ii) BNN designs at this stage are mainly driven by the algorithm community on how to improve the training accuracy; little system and architectural support have been provided on high performance delivery. Due to (i), most existing BNN implementations are realized as hardware accelerators (e.g., through FPGA \cite{nurvitadhi2016accelerating, umuroglu2017finn, zhao2017accelerating, liang2018fp, geng2019lp, geng2019o3bnn}) where the operand bit-width can be flexibly adjusted. Due to (ii), BNN developers are still relying on full-precision software frameworks such as TensorFlow and PyTorch over CPUs and GPUs to emulate the BNN execution. As a result, the lack of architectural \& system support hinders the performance delivery and the general adoption of BNNs.

\begin{table}[!t]
\centering\footnotesize
\caption{Bit-Software-Tensor-Core (BSTC) \cite{li2019bstc} vs. Bit-Tensorcore (BTC).  \emph{uint32} refers to \emph{unsigned int}. \emph{uint64} refers to \emph{unsigned long long int}. \emph{INTU} refer to \emph{integer units}. \emph{SFU} refers to \emph{special function units}.}
\begin{tabular}{|c|c|c|}
\hline
 & \textbf{BSTC}  & \textbf{BTC}   \\ \hline
\textbf{Datatype} & Bit (uint32, uint64) & Bit (uint32) \\ \hline
\textbf{Functionality} & Bit Matrix Multiplication & Bit Matrix Multiplication \\ \hline
\textbf{Tile-A size} & 32$\times$32 or 64$\times$64 & 8$\times$128 \\ \hline
\textbf{Tile-B size} & 32$\times$32 or 64$\times$64 & 128$\times$8 \\ \hline
\textbf{Tile-C size} & 32$\times$32 or 64$\times$64 & 8$\times$8 \\ \hline
\textbf{Hardware units} & INTUs and SFUs & TensorCore Units (TCUs) \\ \hline
\textbf{Processing level} & per warp & per warp \\ \hline
\textbf{GPU Platforms} & Kepler or later ($\ge$CC-3.0) & Turing GPUs ($\ge$CC-7.5) \\ \hline
\end{tabular}
\label{tab:bstc-vs-bhtc}
\end{table}

This situation has been lately changed for GPUs. On the software side, a recent work \cite{li2019bstc} proposed the binarized software tensor core, or BSTC, relying on GPU's low-level hardware intrinsics for efficient 2D bit-block processing, such as \emph{bit matrix multiplication} (BMM) and \emph{bit convolution} (BConv). On the hardware side, the latest NVIDIA Turing GPUs started to support BMM experimentally in their \emph{Tensor Core Units} (TCUs) \cite{turinggpu}. We label this new bit-capability as Bit-Tensor-Core, or BTC. Table~\ref{tab:bstc-vs-bhtc} compares the major features of BSTC and BTC.

In this work, we focus on these bit-tensorcores in Turing GPUs and investigate how they can be fully leveraged for advanced performance delivery for BNNs. This paper thus makes the following major \textbf{contributions:} (i) To the best of our knowledge, this is the first work to investigate this brand new bit-computation capability of GPU tensorcores\footnote{To the best of our knowledge, this feature has not appeared in vendor's library like cuBLAS, cuDNN, TensorRT or other library up to now except \emph{Cutlass} in which it is supported as an experimental, unverified function.}. We designed orchestrated microbenchmarks to investigate the low-level features of BTC. In particular, we observed that the value of stride exhibits considerable performance impact on memory fetch from the global memory via Warp-Matrix-Multiplication-API (WMMA). (ii) Based on our observations, we proposed a new bit data format specially for bit-computation on GPU tensorcores. We showed that without this new bit  format, BTC might not exhibit performance advantage over existing software solutions; (iii) BTC currently only supports \emph{XOR}-based bit-matrix-multiplication. In terms of convolution, traditional approaches \cite{chellapilla2006high, chetlur2014cudnn} that transform convolution to matrix-multiplication fail to work properly for BNNs due to the challenge in padding \cite{electronics8060661}. In this work, we propose a new approach that can fully leverage the bit-computation capability of BTC while effectively resolving this padding issue. We evaluated our design on two NVIDIA Turing GPUs, the results showed that our BTC-based BMM design could bring up to 4.4$\times$ speedup over the vendor's \emph{Cutlass} \cite{cutlass} implementation. Regarding BNN inference performance, compared with state-of-the-art solution \cite{li2019bstc}, our BTC-based approach achieved on average 2.20$\times$ and 2.25$\times$ in latency, and 1.99$\times$ and 1.62$\times$ in throughput for VGG-16 and ResNet-18 on ImageNet. As bit-computation is increasingly common in many HPC and data-analytics scenarios \cite{block2010multi, pedemonte2011bitwise, fusco2013indexing, xu2013bit, ben2016fast, song2017binary}, our techniques can be extended to other applications.

% as well.

%\cite{fang2009frequent, block2010multi, pedemonte2011bitwise, fusco2013indexing, xu2013bit, ben2016fast, song2017binary, ashkiani2018dynamic}

\begin{figure}[!t]
\centering
\vspace{-8pt}
\includegraphics[width=0.8\columnwidth]{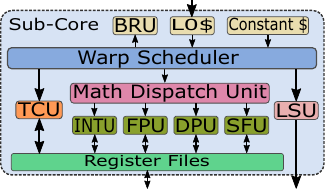} 
\caption{A subcore of a Turing GPU SM. \emph{BRU} is branch unit. \emph{\$} refers to cache. LSU is the load-store-unit. INTU is the integer-unit. FPU is the floating-point-unit. DPU is the double-precision-unit. SFU is the special-function-unit \cite{li2016sfu}. TCU is the tensorcore-unit, which has its independent data path.}
\label{fig:tensorcore}
\end{figure}

\section{Related Work}

We focus on the performance issues of BNN implementation and the GPU tensorcores in this section. Regarding the algorithm design for BNNs, please refer to this survey \cite{electronics8060661}. 

\vspace{4pt}\noindent \textbf{BNN Implementation} The major purpose of BNN implementation is to leverage the system and architectural features of the platforms to satisfy the stringent latency and throughput constraints when deploying BNNs in HPC, cloud and embedded applications, while reducing the area and energy cost \cite{nurvitadhi2016accelerating, umuroglu2017finn, zhao2017accelerating, mcdanel2017embedded, liang2018fp, hubitflow2018, geng2019lp, geng2019o3bnn, geng2020o3bnn}. Most of these implementations focus on FPGA \cite{nurvitadhi2016accelerating, umuroglu2017finn, zhao2017accelerating, liang2018fp, geng2019lp, geng2020o3bnn} due to FPGA's design flexibility at the bit level. Regarding general-purpose platforms, an existing CPU work \cite{hubitflow2018} relies on the AVX/SSE vector instructions to derive good bit computation performance. It focuses on BMM and transforms Bconv to BMM through \emph{im2col()} with costly pre- and post-processing. Another evaluation work \cite{lin2017towards} compares CPU, GPU, FPGA and ASIC based BNN designs, clarifying that the major performance restriction of CPUs and GPUs is \emph{the extremely low utilization due to the challenge in extracting fine-grained parallelism}. Noticeably, the reported GPU utilization is $1\%$ only \cite{lin2017towards}. To improve GPU utilization and extract bit-level-parallelism, a recent work \cite{li2019bstc} proposed the binarized-soft-tensor-core on top of GPU's SMs and leverages low-level hardware intrinsics for harvesting the bit capability of GPUs. For BSTC, the performance gains from better utilization of the conventional integer/logic units (i.e., INTUs and SFUs, see Figure~\ref{fig:tensorcore}). This work is different as we focus on the brand new bit computation capability of the latest Turing TCUs, and showcase how to gain the most performance from this new functional units.

\vspace{2pt}\noindent \textbf{GPU Tensorcore} Driven by the demand of training large-scale DNNs, designing specialized low-precision dense matrix multiplication accelerators has become a popular trend. Particularly, Google presented \emph{Tensor-Processing-Units} \cite{jouppi2017datacenter}; Intel announced the Nervana \emph{Neural-Network-Processors}  for tensor operations; NVIDIA integrated the \emph{Tensorcore Units} into their Volta and Turing GPUs; Qualcomm included the \emph{Hexagon-Tensor-Accelerator} into their Hexagon 855 chip. 

This work focuses on the tensorcores of GPUs (see Figure~\ref{fig:tensorcore}). Since being firstly introduced in the Volta architecture \cite{nvidia2018volta}, the tensorcore becomes one of the spot-light for GPGPU research. The relevant works can be summarized in two categories: \textbf{(a) Characterization}. In \cite{jia2018dissecting} and \cite{jia2019dissecting}, Jia et al. dissected the Volta V100 and the Turing T4 GPUs through microbenchmarking. They depicted the detailed mapping mechanism from elements of a matrix tile to registers of a warp-lane in the HMMA instructions for FP16 matrix multiplication. They found that the 32 threads of a warp are essentially divided into 8 thread groups, where the 4 threads per group cooperatively work on the same regions of matrix $C$ by fetching elements from different parts of matrix $A$ and matrix $B$. Markidis et al. \cite{markidis2018nvidia} studied the programmability, performance and precision of the Volta tensorcores and proposed a technique to compensate the accuracy loss due to precision degradation from FP32 to FP16. Raihan et al. \cite{raihan2019modeling} investigated the design details of the tensorcores in Volta and Turing GPUs and built an architecture model for the tensorcores in GPGPU-Sim. They characterized the WMMA APIs and clarified how the operand sub-matrix elements were mapped for FP16 GEMM in Volta tensorcores, and FP16/Int8/Int4 GEMM in Turing tensorcores. However, they did not investigate the 1-bit computation mode. Hickmann and Bradford \cite{hickmann2019experimental} proposed a testing method for assessing the compliance of IEEE standard, hardware microarchitecture, and internal precision of the Volta tensorcores. \textbf{(b) Application}. Haidar et al. \cite{haidar2018harnessing} proposed a mixed-precision iterative refinement method to approach FP64 precision using FP16-based GPU tensorcores in LU factorization, acting as the first effort to apply GPU tensorcores for non-machine-learning applications. Sorna et al. \cite{sorna2018optimizing} applied FP16 tensorcores for FFT acceleration. Blanchard et al. \cite{blanchard2019mixed} thoroughly analyzed the rounding error of matrix multiplication and LU factorization when using the tensorcores. Dakkak et al. \cite{dakkak2019accelerating} showed that the tensorcores, which were originally designed for 2D FP16 GEMM, can be adopted for 1D array reduction and scan. 

%Zhang and Wu \cite{zhang2019high} proposed a QR factorization algorithm accelerated by the tensorcores and attained considerable speedups over cuSOLVER.

Most of these works, however, focused on FP16 mixed-precision matrix-multiply in Volta tensorcores. They either evaluated their performance, programmability, accuracy, hardware design, or looked into alternative applications other than GEMM, aiming to preserve higher precision. None of them have investigated the latest bit computation capability of the GPU tensorcores. In addition, no existing works, to the best of our knowledge, have ever reported the potential performance impact from the stride of segmented memory load, and how to circumvent the challenges in accelerating convolutions through the tensorcores. Furthermore, until writing the paper, we have not seen any works leverage GPU tensorcores for the acceleration of BNNs.

\section{GPU Bit Tensorcores}

%In this section, we briefly introduce the GPU tensorcores. 

%We then characterize the \texttt{popc-xor} bit computation capability of the Turing tensorcores in the next section.

\subsection{GPU Tensorcores}

Since the Volta architecture (CC-7.0), NVIDIA GPUs have introduced a novel type of function units known as \emph{Tensorcores} into the streaming multiprocessors (SMs) for accelerating low-precision general matrix multiplication (GEMM). In Volta, each tensorcore processes 64 FP16 FMA operations per cycle \cite{nvidia2017v100}. The only supported datatype for Volta tensorcores is FP16. For Turing (CC-7.5), more  datatypes are supported, including FP16, signed/unsigned int-8, int-4, and recently a bit as well. Please refer to \cite{nvidia2017v100, jia2018dissecting, raihan2019modeling} for more hardware details about Volta and Turing TCUs.

\subsection{CUDA WMMA}

Since CUDA Runtime-9.0, the  \emph{Warp Matrix Multiplication API} (WMMA) has been introduced for operating the tensorcores in Volta and Turing GPUs. The idea is to partition the three input and one output matrices into tiles, where each warp processes the multiplication of one tile ($T_D=T_A\times T_B + T_C$). WMMA provides the necessary primitives to operate on the bit-tiles (e.g., loading input tiles, tiled multiplication, storing output tile): \emph{load\_matrix\_sync}, \emph{mma\_sync}, \emph{store\_matrix\_sync}. These primitives are executed by the 32 threads of a warp cooperatively. For FP16, the tile is further partitioned into 32 fragments while each thread fetches a fragment of data into its register files. Although the vendor's official documents have not revealed the exact mapping schemes, existing works have figured them out through microbenchmarking \cite{jia2018dissecting, raihan2019modeling}.

\subsection{Cutlass Library}

Currently the vendor's high-performance linear-algebra library \emph{cuBLAS} has not supported BMM on Turing tensorcores. However, their open-source GEMM library -- \emph{Cutlass} \cite{cutlass} has integrated it as an experimental and non-verified feature. BMM is realized using the WMMA API. The input matrix $A$ is in row-major bit format (compacted as 32-bit unsigned int), $B$ is in column-major bit format (also compacted as 32-bit unsigned int). The accumulated input matrix $C$ and the result matrix $D$ are in row-major 32-bit signed int format. $C$ and $D$ are usually the same matrix. BMM in \texttt{Cutlass} conducts 0/1 dot-product while BNN demands +1/-1 dot-product, as will be discussed later.

% whereas BMM in BNN demands +1/-1 dot-product. 

%We discuss this in more detail later. In our evaluation, we compare the performance of our BMM design with \texttt{Cutlass}.

\section{BTC Characterization}

\begin{figure}[!t]
\begin{lstlisting}[caption=WMMA 1-bit definition and operations in \emph{crt/mma.h}, captionpos='b', label={lst:header}]
namespace experimental { 
  namespace precision { struct b1; } // define 1-bit datatype
  enum bmmaBitOp { bmmaBitOpXOR = 1 }; //only XOR is supported as the mul-op for BTC
  enum bmmaAccumulateOp { bmmaAccumulateOpPOPC = 1 }; } //popc acts as the acc-op for BTC
\end{lstlisting}
\end{figure}

To operate on the bit datatype for Turing GPUs, CUDA WMMA defines the 1-bit precision and the bit operations in an independent \emph{"experimental"} namespace, as listed in Listing~\ref{lst:header}. \texttt{XOR} and \texttt{POPC} for bits (+1/-1) correspond to multiply and accumulate for floating-point/integer datatypes.

\begin{figure}[!t]
\vspace{-14pt}
\begin{lstlisting}[caption=PTX code for Bit-Matrix-Multiply-API (BMMA), captionpos='b', label={lst:ptx}]
wmma.load.a.sync.aligned.row.$m8n8k128$.global.b1 {%r21}, [%rd8], %r20; //load_matrix_sync()
wmma.load.b.sync.aligned.col.$m8n8k128$.global.b1 {%r22}, [%rd5], %r20; //load_matrix_sync()
wmma.load.c.sync.aligned.row.$m8n8k128$.global.s32 [%rd30],{%r55,%r64},%r60;//load_matrix_sync()
wmma.mma.$xor.popc$.sync.aligned.row.col.$m8n8k128$.s32.b1.b1.s32 {%r23,%r24}, {%r21}, {%r22},  {%r19,%r19}; //bmma_sync()
wmma.store.d.sync.aligned.row.$m8n8k128$.global.s32 [%rd34], {%r63,%r64},%r60;//store_matrix_sync()
\end{lstlisting}
\end{figure}

Five APIs are provided for loading the bit-tile \emph{A}, the bit-tile \emph{B}, the int-tile \emph{C}, and storing the int-tile \emph{D}, as well as the multiplication: $D=C+A\times B$. For the bit-matrix-multiply-API (BMMA), only a single computation paradigm is defined: the bit-tile A is in row-major of size (8, 128); the bit-tile B is in column-major of size (128, 8); the int-tile C and D are square matrices in row-/column-major of size (8, 8). The bit-tile \emph{A} and \emph{B} are compacted as 32 unsigned ints, each with 32 bits. Therefore, the bit-tile \emph{A} and \emph{B} each occupies 128 bytes. The int-tile \emph{C} and \emph{D} each occupies 8$\times$8$\times$4=256 bytes. Listing~\ref{lst:ptx} shows the  \emph{Parallel Thread Execution} (PTX) -- the low-level GPU virtual machine ISA code for the five BMMA APIs. The shape qualifier "\texttt{m8n8k128}" in Line~2-6 indicate that the bit-tile-multiplication processed per warp is in size (8,128)$\times$(128,8)=(8,8). "\texttt{sync}" means the instruction wait for all warp lanes to synchronize \cite{li2015fine} before proceeding. The "\texttt{layout}" qualifier specifies if the tile is stored with a row-major or column-major order in memory. "\texttt{type}" indicates the precision of the tile. Using \texttt{int32} for tile-\emph{C} and \emph{D} is to avoid potential overflow during the accumulation. For matrix-multiplication, tile-C and D are usually in the same size. Thus, the five APIs can be grouped into three set: \emph{load}, \emph{store},  and \emph{computation}. We investigate each of them in the following to figure out potential design guidelines. Regarding the hardware platform, see Section~7 and Table~\ref{tab:gpu}.

\subsection{BMMA Load}

We first concentrate on \emph{bmma\_load}, as memory load is the most crucial factor for GEMM on GPU \cite{tan2011fast}. The load API is
\begin{lstlisting}[basicstyle={\tiny\sffamily\bfseries}]
  void load_matrix_sync (fragment<...> &tileA, const T* mptr, unsigned ldm, layout_t layout);
\end{lstlisting}
It waits for all the threads of a warp to arrive, and then loads a bit tile (i.e, a matrix fragment) from the device memory. It has three parameters. "\emph{mptr}" is a 256-bit aligned pointer pointing to the first element of the matrix in memory. The memory here can be global or shared memory. "\emph{layout}" can be row- or column-major, but for BMMA, there is only a single choice --- \emph{mem\_row\_major} for matrix-A and \emph{mem\_col\_major} for matrix-B. "\emph{ldm}" is the stride in element between consecutive rows (in row major) or columns (in column major) and must be a multiple of 16 bytes), according to \cite{cuda2012programming}. We find that for shared memory, this is the case; but for global memory, a multiple of 32 is also feasible (despite with unpredicted results). To see the impact of \emph{mptr} (i.e., memory type) and \emph{ldm} on the performance of the load primitive, we measure its average per-thread latency using the \texttt{clock()} instruction. We add a memory fence operation before the measurement to ensure that the data fetching has finished.

\begin{figure*}[!htb]
\minipage{0.5\columnwidth}
\includegraphics[width=1.05\columnwidth]{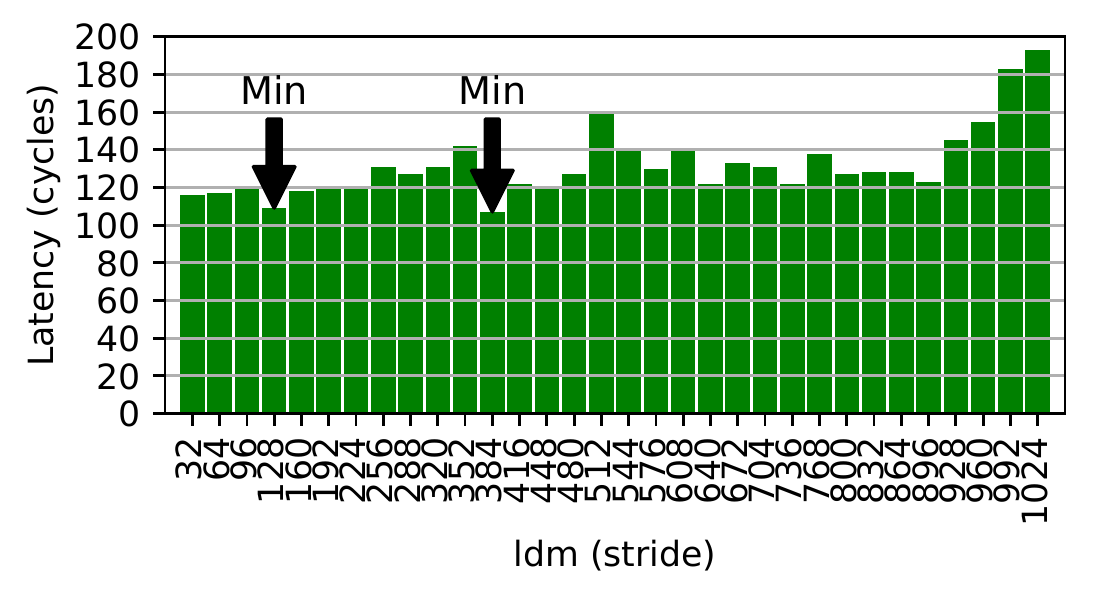} 
\caption{RTX2080 global mem load }
\label{fig:load_global_2080}
\endminipage\hfill
\minipage{0.5\columnwidth}
\includegraphics[width=1.05\columnwidth]{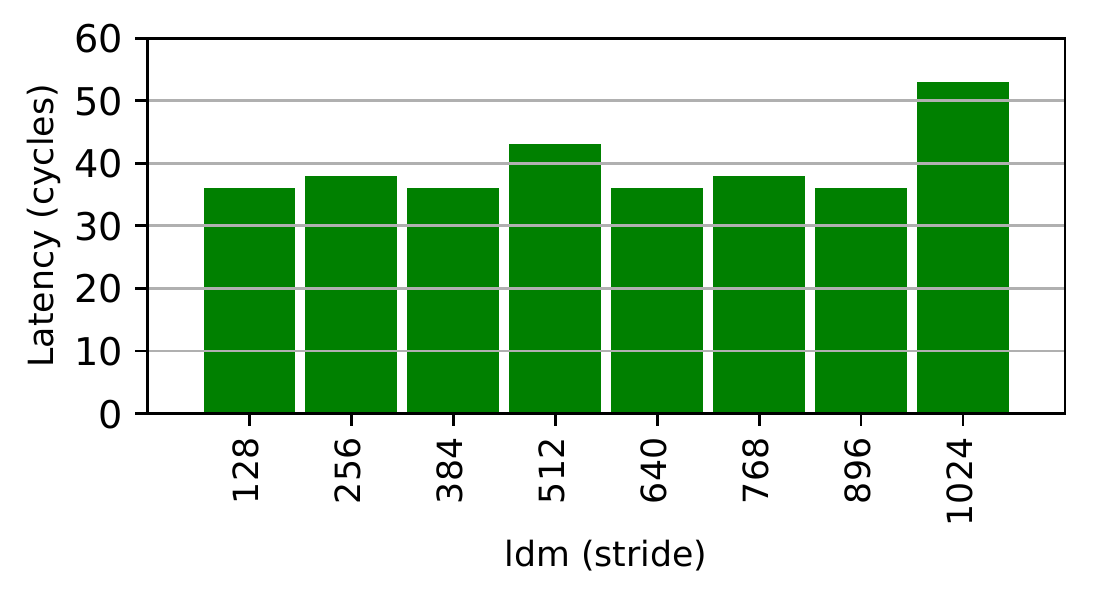} 
\caption{RTX2080 shared mem load }
\label{fig:load_shared_2080}
\endminipage\hfill
\minipage{0.5\columnwidth}
\includegraphics[width=1.05\columnwidth]{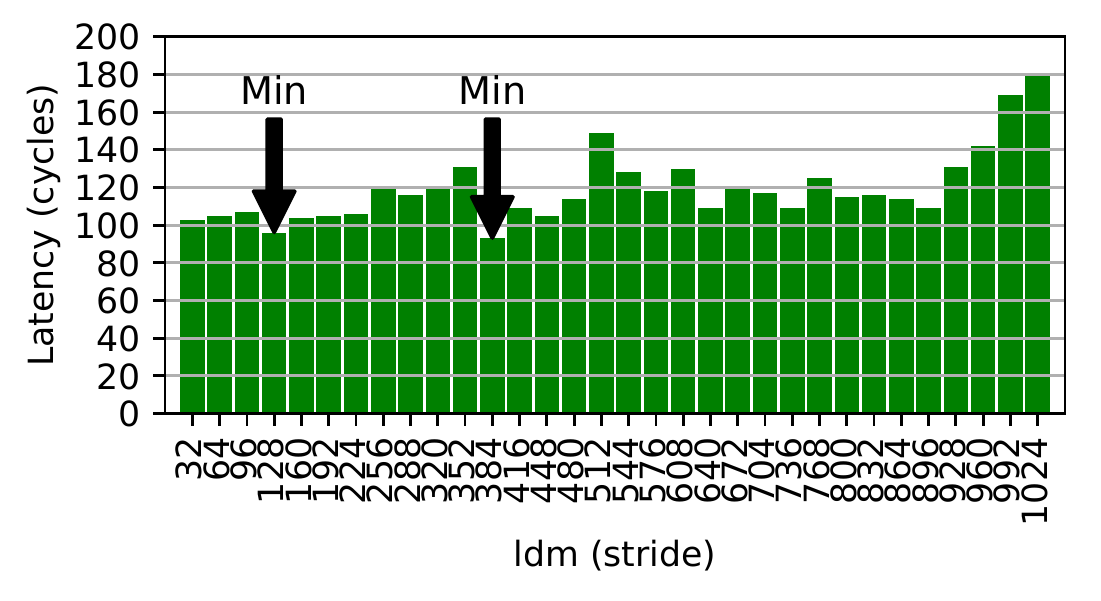} 
\caption{RTX2080Ti global mem load }
\label{fig:load_global_2080ti}
\endminipage\hfill
\minipage{0.5\columnwidth}
\includegraphics[width=1.05\columnwidth]{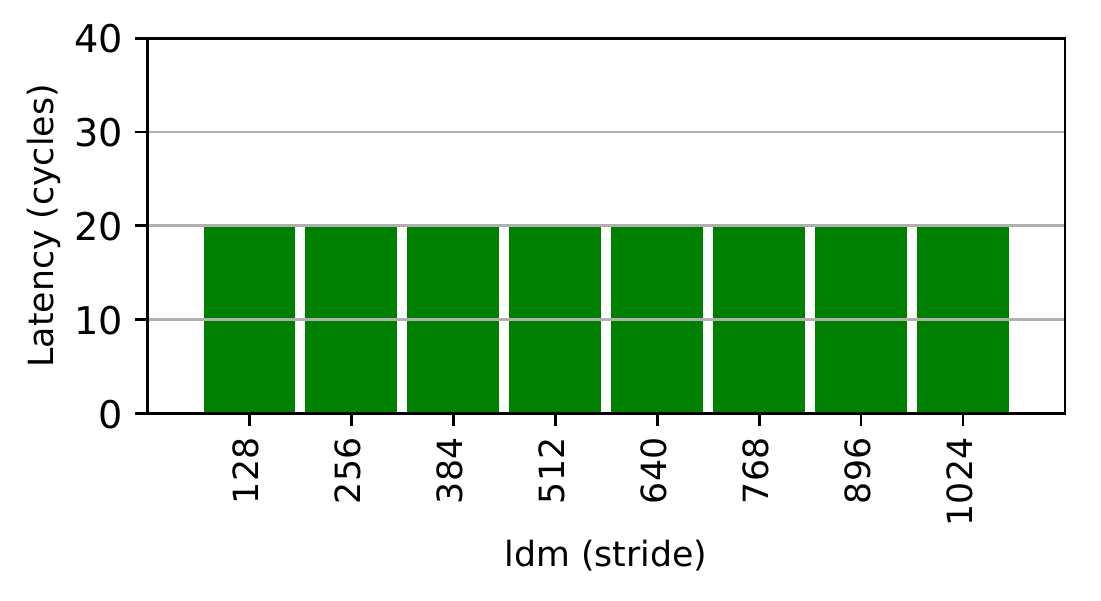} 
\caption{RTX2080Ti shared mem load }
\label{fig:load_shared_2080ti}
\endminipage
\end{figure*}

\begin{figure*}[!htb]
\minipage{0.5\columnwidth}
\includegraphics[width=1.05\columnwidth]{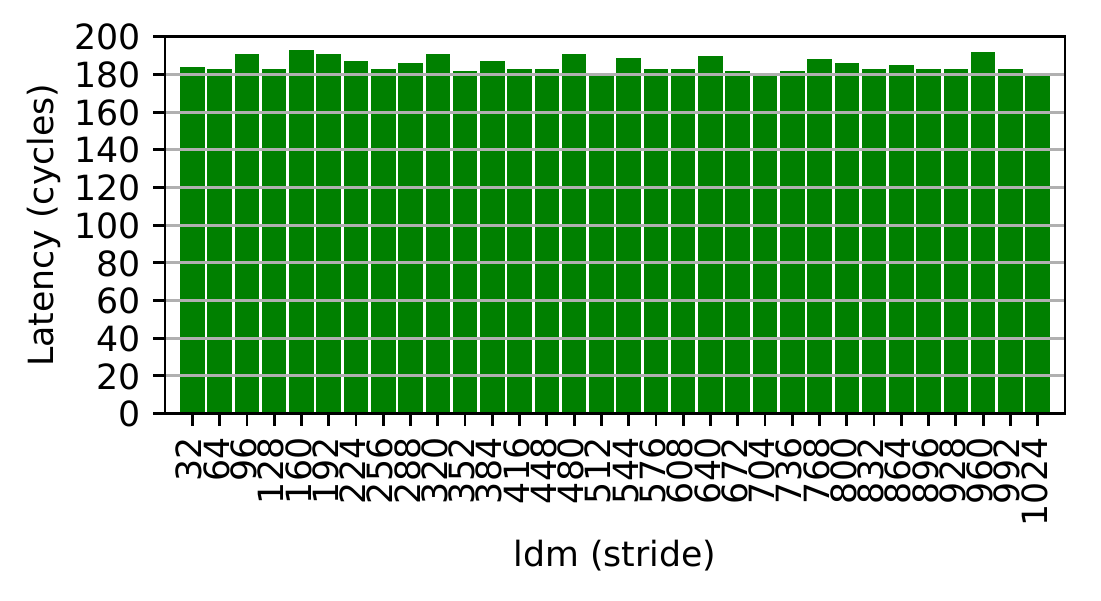} 
\caption{RTX2080 global mem store }
\label{fig:store_global_2080}
\endminipage\hfill
\minipage{0.5\columnwidth}
\includegraphics[width=1.05\columnwidth]{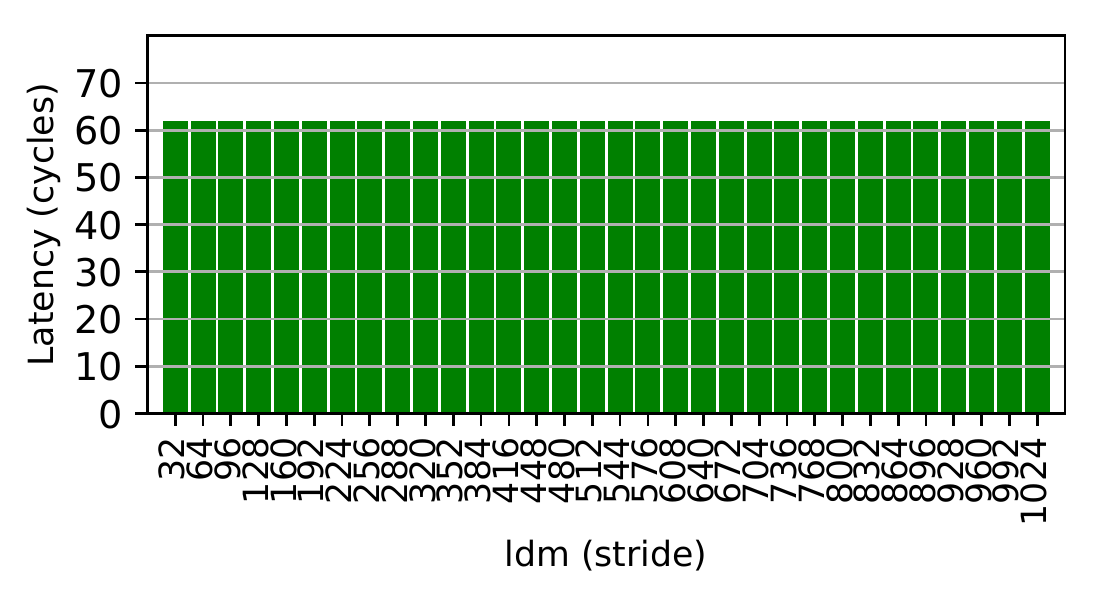} 
\caption{RTX2080 shared mem store }
\label{fig:store_shared_2080}
\endminipage\hfill
\minipage{0.5\columnwidth}
\includegraphics[width=1.05\columnwidth]{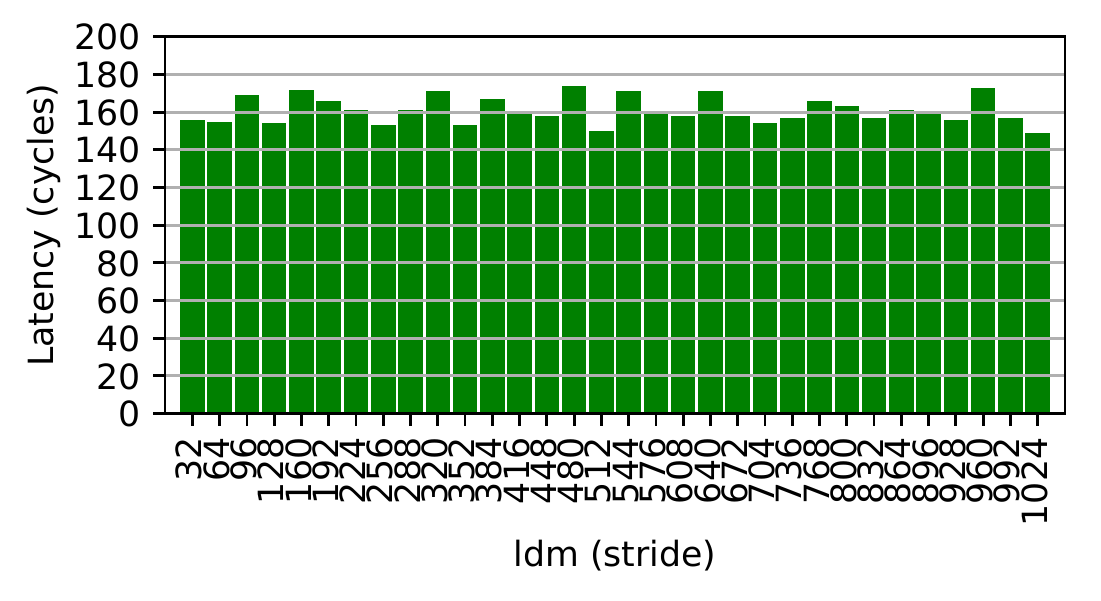} 
\caption{RTX2080Ti global mem store }
\label{fig:store_global_2080ti}
\endminipage\hfill
\minipage{0.5\columnwidth}
\includegraphics[width=1.05\columnwidth]{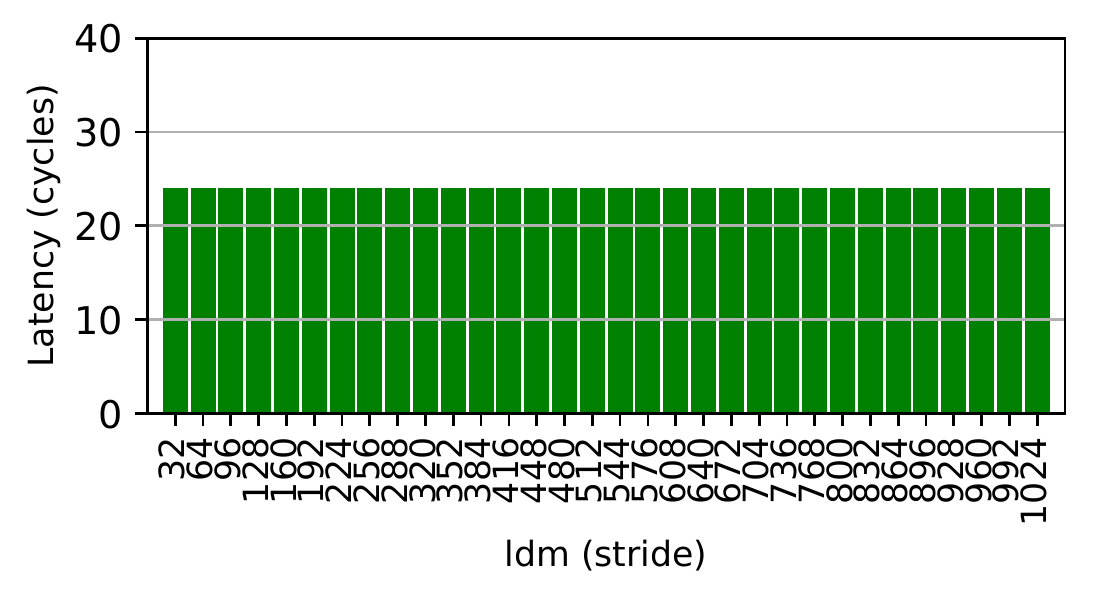} 
\caption{RTX2080Ti shared mem store }
\label{fig:store_shared_2080ti}
\endminipage
\end{figure*}

\begin{figure*}[!htb]
\minipage{0.49\columnwidth}
\includegraphics[width=1.1\columnwidth]{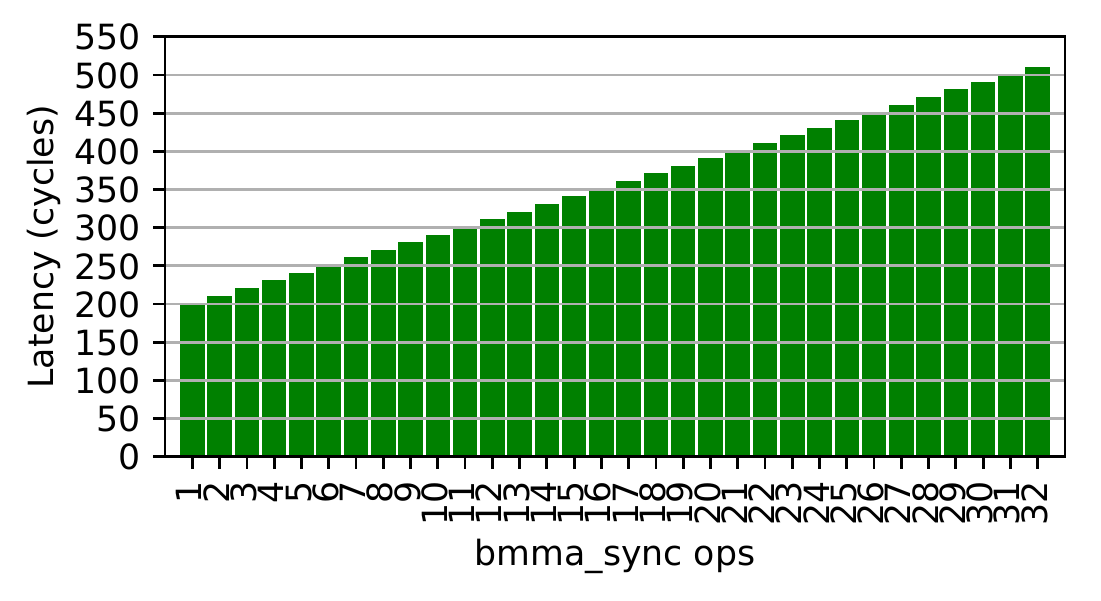} 
\caption{RTX2080 bmma w/ same C}
\label{fig:bmma_sameC_rtx2080}
\endminipage\hfill
\minipage{0.49\columnwidth}
\includegraphics[width=1.1\columnwidth]{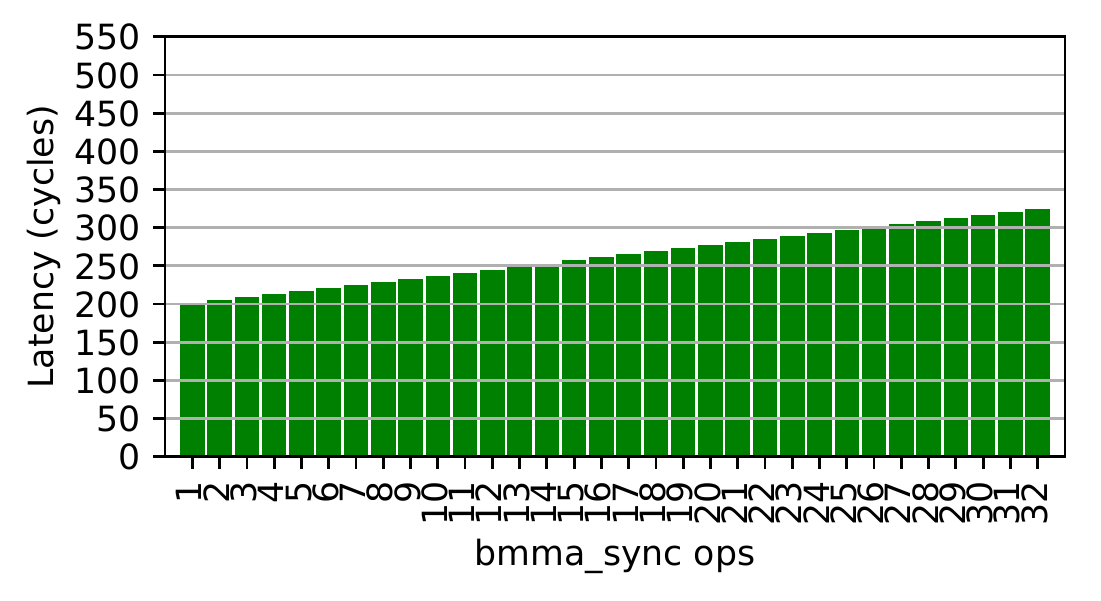} 
\caption{RTX2080 bmma w/ diff C}
\label{fig:bmma_diffC_rtx2080}
\endminipage\hfill
\minipage{0.51\columnwidth}
\includegraphics[width=1.1\columnwidth]{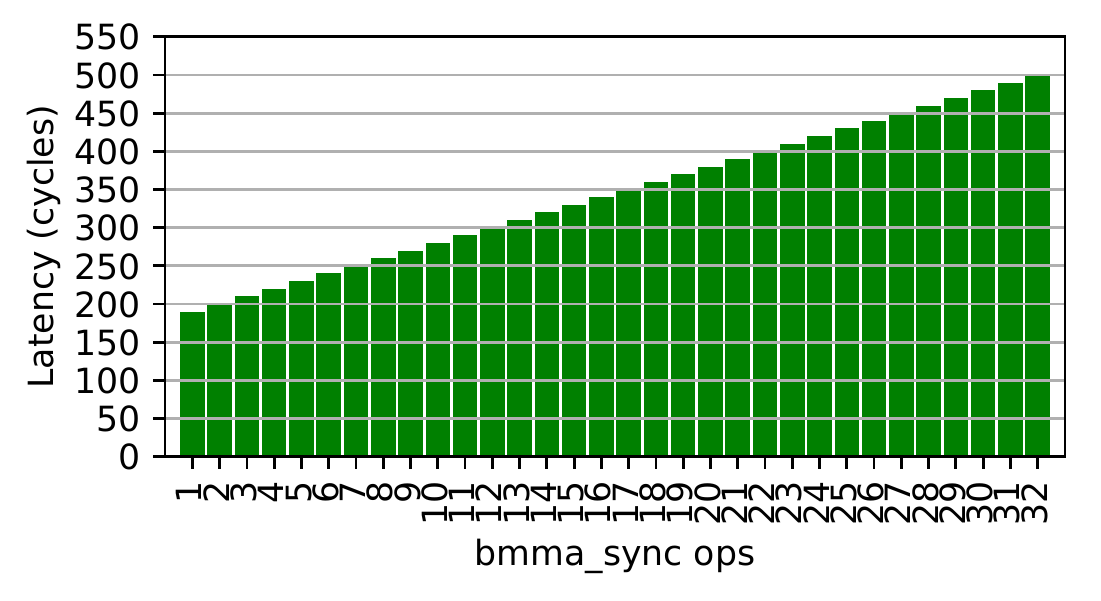} 
\caption{RTX2080Ti bmma w/ same C}
\label{fig:bmma_sameC_rtx2080ti}
\endminipage\hfill
\minipage{0.49\columnwidth}
\includegraphics[width=1.1\columnwidth]{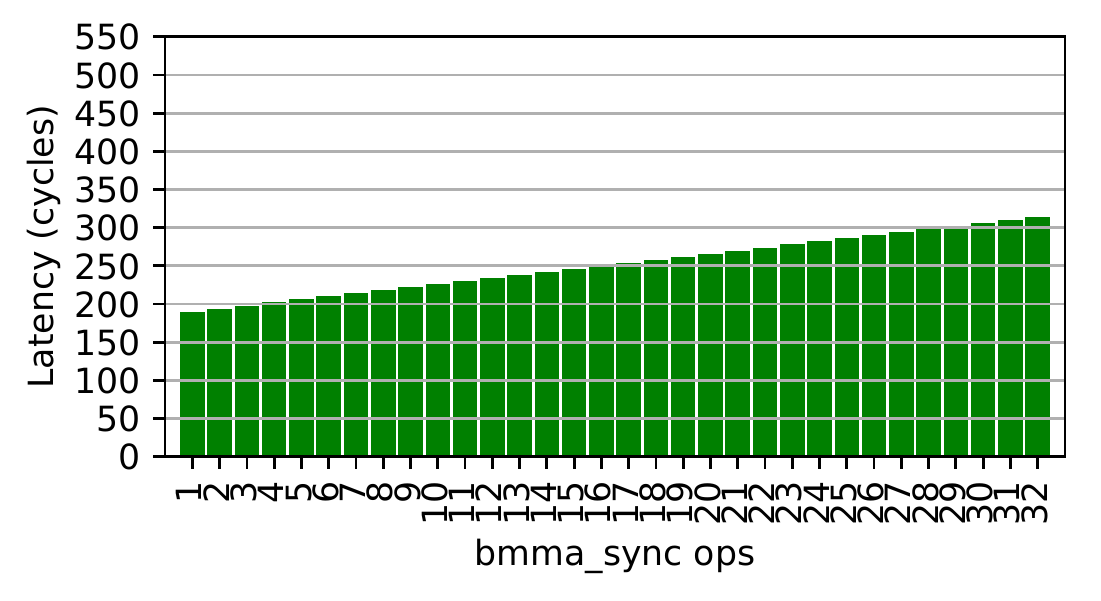} 
\caption{RTX2080Ti bmma w/ diff C }
\label{fig:bmma_diffC_rtx2080ti}
\endminipage
\end{figure*}

Figure~\ref{fig:load_global_2080}, \ref{fig:load_shared_2080}, \ref{fig:load_global_2080ti}, \ref{fig:load_shared_2080ti} show the average latency with respect to different values of \emph{ldm} for \emph{load\_matrix\_sync()} on global and shared memory of RTX-2080 and RTX-2080Ti GPUs. As \emph{ldm} is the stride between consecutive rows of the matrix, it should be application dependent (e.g., for a 1024$\times$1024$\times$1024 BMM, \emph{ldm} should be 1024) and the raw latency should be irrelevant to \emph{ldm}. However, counterintuitively it has a strong impact on the performance of fetching a bit-tile from the global memory. As can be seen in Figure~\ref{fig:load_global_2080} and Figure~\ref{fig:load_global_2080ti}, \emph{ldm=128} and \emph{384} exhibit the shortest latency. Regarding shared memory, (1) accessing shared memory shows more than 5$\times$ less latency than global memory; (2) the latency for RTX2080Ti shared memory access is less than RTX2080, and is unchanged with \emph{ldm}. 

We then consider "\emph{tileA}", and see how the bit-tile ($128\times8$ bits) is distributed among the lanes of a warp. Similar to \cite{raihan2019modeling}, we let each lane print out the value of data it fetches. Based on the value, we can identify the mapping mechanism. We find that, similar to FP16 and Int8, lanes in BMMA also establish 8 thread groups --- 4 consecutive lanes per group. Each thread group corresponds to a 128 bit row. Within a 128 bit row, each lane accounts for a 32-bit portion (4 bytes). This partially explains why \emph{ldm=128} delivers the shortest latency: the 32 lanes of the warp constitute a coalesced memory access, where the 32 4-byte access are merged as a single memory request. Regarding why \emph{ldm=384} also exhibits good performance, we suspect this might be because the Turing L1 data cache is essentially partitioned into two sectors with independent ports, similar to the L1/texture cache in Maxwell and Pascal GPUs \cite{li2017locality}. It conserves the data in an interleaving way at a step of 32 bytes. Consequently, \emph{ldm=256} (32B) may trigger a sector-port conflict for simultaneous memory fetches from the same warp but \emph{ldm=384} may not. This is confirmed by the observation that \emph{ldm=128+256X} (e.g., 384, 640, 896) all demonstrate relatively low latency in Figure~\ref{fig:load_global_2080} and \ref{fig:load_shared_2080ti}.

%outstanding performance is due to good locality exploitation at the L1 cache but meanwhile avoiding the L1 cache bank conflict. As we know the L1 cache and shared memory in Turing GPUs are sharing the same SRAM storage, the L1 cache should also be organized with 32 banks. Given a warp in BMMA load is separated into 8 groups, the 16 bytes per group form a \emph{LDG.128} memory transaction, 

\subsection{BMMA Store} 
The store operation is different from the load operation in that each element is a 32-bit signed integer. The store API is:
\begin{lstlisting}[basicstyle={\tiny\sffamily\bfseries}]
  void store_matrix_sync (T* mptr, const fragment<...> &tileC,  unsigned ldm, layout_t layout);
\end{lstlisting}
Again, it waits until all warp lanes arrived before storing \emph{tileC} into memory. "\emph{mptr}" must be a 256-bit aligned pointer referring to the first element. "\emph{ldm}" describes the stride in elements between consequent rows in C, and must be a multiple of 16 bytes (with integer, it corresponds to 4 elements). "\emph{layout}" can be row-major or column-major. 

We measure the average latency with respect to the stride \emph{ldm} on global and shared memory of RTX-2080 and RTX-2080Ti GPUs, as shown in Figure~\ref{fig:store_global_2080}, \ref{fig:store_shared_2080}, \ref{fig:store_global_2080ti}, \ref{fig:store_shared_2080ti}, respectively. Unlike load, the latency histograms for store do not exhibit obvious patterns. We also attempt to investigate how the resulting int-tile \emph{tileC} is distributed among the lanes. Our findings show that: (i) If it is row-major, then within the $8\times8$ int tile, each two consecutive elements (from a row) are stored in two adjacent registers of a lane. For example, suppose the $8\times8$ elements are E0 to E63 and each lane uses R4 and R5 to store the integer tile, then (E0, E1) are stored in R4 and R5 of lane-0, (E2, E3) are stored in R4 and R5 of lane-1, and so on. (ii) If it is column-major, then each two consecutive elements (from a column) are stored in two adjacent registers of a lane (i.e., transposed from the row-major layout). When storing, the two adjacent registers are encoded as one \texttt{STG.E.64} memory store for the entire warp, as if storing an FP64 data.

\subsection{BMMA Computation}

Finally, we discuss the bit-matrix-multiply API: 
\begin{lstlisting}[basicstyle={\tiny\sffamily\bfseries}]
 void bmma_sync(fragment<...> &tileD, const fragment<...> &tileA, 
   const fragment<...> &tileB, const fragment<...> &tileC, 
   experimental::bmmaBitOp=experimental::bmmaBitOpXOR,
   experimental::bmmaAccumulateOp=experimental::bmmaAccumulateOpPOPC);
\end{lstlisting}
It waits until all lanes are available for conducting the BMMA operation: \emph{tileD = POPC(tileA XOR tileB) + tileC}. Unlike the condition for FP16 and Int8 where a group of SASS assembly operations are generated \cite{jia2018dissecting, raihan2019modeling, jia2019dissecting}, \emph{bmma\_sync} is only translated into a single SASS code:  
\begin{lstlisting}[basicstyle={\tiny\sffamily\bfseries}]
  BMMA.88128.$XOR$.$POPC$ 		R2,     R8.ROW,    R9.COL,   R2 ; 
\end{lstlisting}
Our idea here is to measure its raw latency and estimate how much parallelism, including warp-level-parallelism (WLP) and instruction-level-parallelism (ILP), are required to saturate the tensorcore pipeline and hide the latency.

Figure~\ref{fig:bmma_sameC_rtx2080}, \ref{fig:bmma_diffC_rtx2080}, \ref{fig:bmma_sameC_rtx2080ti}, \ref{fig:bmma_diffC_rtx2080ti} illustrate the total latency of increasing the number of repeated \emph{bmma\_sync} operations for the same \emph{tileC/tileD}, and different \emph{tileC/tileD} on the two GPUs. The raw latency of \emph{bmma\_sync} is $\sim$201 cycles on RTX2080 and $\sim$190 cycles on RTX2080Ti. As shown in the figure, the incremental latency with each one more \emph{bmma\_sync} operation is 10 cycles when \emph{tileC} \& \emph{tileD} are identical for all operations, and is 4 cycles when \emph{tileC} \& \emph{tileD} are different on both platforms. This implies that the pipeline stage delay is around 4 cycles. When using the same accumulator, 6 extra cycles are needed. Given the raw latency of $\sim$200 cycles, and the fact that Turing GPU SM comprises four sub-cores (each subcore can issue one instruction per cycle), with at maximum 32 warps per SM for Turing (so WLP=32), we roughly require ILP=200/4$\times$4/32$\approx$7 to saturate the entire tensorcore pipeline. In other words, with 8 independent \emph{bmma\_sync} operations, we should approach the theoretical computation bandwidth of the tensorcores.

\section{BMM and BConv with BTC}
We present our designs for BTC-based BMM and BConv, which are the core functions for the fully-connected layer and convolution layer of BNNs.

\subsection{FSB Data Format for BTC}

\begin{figure}[!t]
\includegraphics[width=\columnwidth]{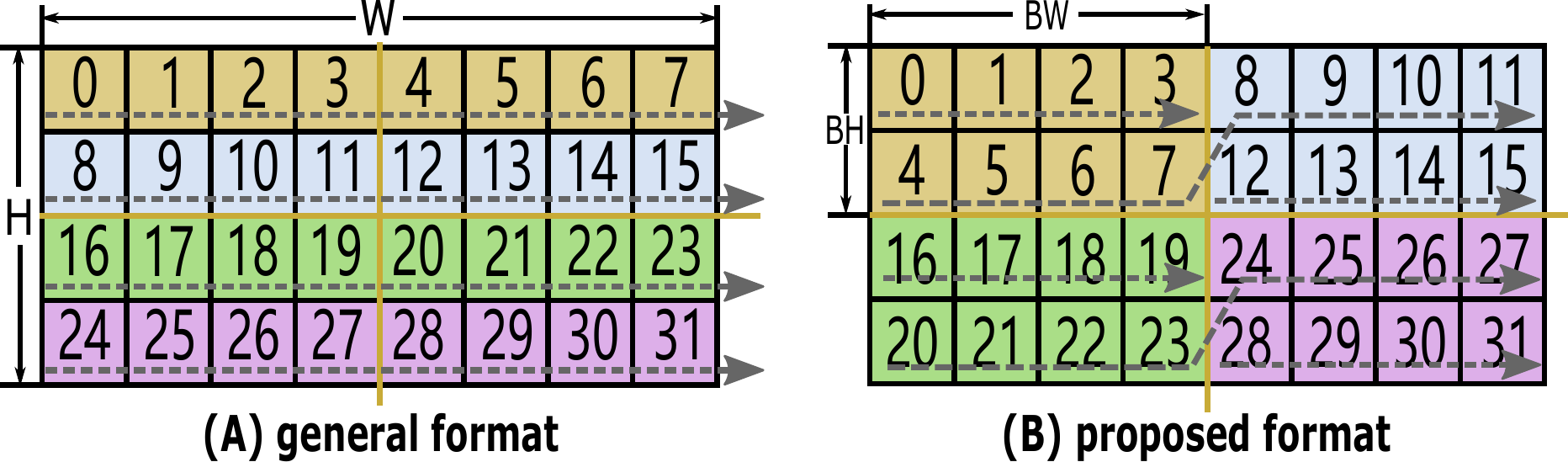} 
\caption{Fixed-Stride-Bit (FSB) format based on a tile of BH$\times$BW.}
\label{fig:newformat}
\end{figure}

In Section~4.1, we have observed that the value of \emph{ldm} can strongly affect the performance of \texttt{load\_matrix\_sync} from global memory, where \emph{ldm}=128 and 384 exhibit the best performance. Our idea thus is whether we can essentially fix the value of \emph{ldm} firmly to 128 or 384. As a result, rather than storing the bits completely sequential and using the matrix width for \emph{ldm}, as practiced by the \emph{Cutlass} library and suggested by CUDA programming guide, we propose a new 2D bit data format where bits are stored in a unit of 128$\times$8 bit-tile. An analogous example is shown in Figure~\ref{fig:newformat}. From the 1D general format to the 2D new format, an array of 8$\times$4 bits (H=4, W=8) is converted with a tile size of 4$\times$2 (BH=2, BW=4). For BTC, since 384 is not a power of 2, dividing 384 may incur troublesome reminder handling, we thus use 128 as BW and 8 as BH for the new format. If the original bits are organized in row-major, both the in-tile and tile-wise order of the new format are in row-major (as the case in Figure~\ref{fig:newformat}); otherwise, both are organized in column-major. Since the new format only changes the way how bits are stored and fetched, no extra space is needed. However, if the width of the original matrix (i.e., W) can not be divided by 128 (i.e., BW), for the convenience of index calculation, we pad the row to be a factor of 128, which may occupy some extra space. Note, in order to load via \emph{load\_matrix\_sync()}, such a kind of padding is required anyway. Similar requirement has been imposed by \emph{pitch} in \emph{cudaMemcpy2D()}. The temporal overhead only occurs at array index calculation, which is almost negligible.

\subsection{BMM for FC Layer}

BMM in BNN is different from GEMM because: (a) \emph{Input}. The elements of matrix-A and B are binary values: +1 and -1. A normal floating-point or int number is binarized via:
\begin{equation}
\label{eq:binarize}
 x^b = sign(x) = 
 \begin{cases}
   \ 1\quad & \text{if}\ x\ge0  \\
  -1\quad &\text{otherwise}
 \end{cases}
\end{equation}
In an FC layer, both A and B have to be binarized ahead of BMM. However, the binarization of B (i.e., weights) can be performed offline after the training; only the binarization of A is in the critical path of inference. Existing work has shown that such a binarization can be achieved efficiently through the \texttt{\_\_ballot()} function of GPUs \cite{li2019bstc}; (b) \emph{Computation}. The dot-product of GEMM is $y_{i,j}=\sum_{k=0}^{n-1}a_{i,k}b_{k,j}$ where $n$ is the vector length. In terms of BMM, as $a$ and $b$ become bit-vectors, if using bit-1 to denote +1, and bit-0 to denote -1, it can be shown that the $\pm1$ dot-product becomes:
\begin{equation}
\label{eq:dot-product}
 v = \vec{a}\cdot\vec{b} = n-2\times popc(\vec{a}\ xor \ \vec{b}) = 2\times popc(\vec{a}\ xnor \ \vec{b})-n
\end{equation}
where $n$ is the bit-vector length. $xor$ and $xnor$ are logical exclusive-or and exclusive-nor. The $xnor$ expression has widely been used for BNN algorithm research \cite{courbariaux2016binarized, rastegari2016xnor} and FPGA/ASIC implementation \cite{umuroglu2017finn, nurvitadhi2016accelerating} while GPU tensorcores currently only support $xor$ for BMM computation. $popc$ stands for population count, which counts the number of bit-1s in the bit vector. (c) \emph{Output}. The elements of the output matrix-C are full-precision integer values. However, in an FC layer, it can be binarized after a threshold operation (discussed later), reducing memory access. Therefore, the third difference with GEMM is that the output-C can be binarized before the store.

\begin{figure}[!t]
\begin{lstlisting}[caption=BMM baseline implementation, captionpos='b', label={lst:bmm1}]
__global__ void BMM(unsigned *A, unsigned *B, int *C, int A_height, int A_width, int B_width) {
  using namespace nvcuda::wmma::experimental;
  int bx = blockIdx.x*blockDim.y+threadIdx.y;  int by = blockIdx.y;
  wmma::fragment<wmma::matrix_a,8,8,128, precision::b1,wmma::row_major>a_frag;//tile A
  wmma::fragment<wmma::matrix_b,8,8,128, precision::b1,wmma::col_major>b_frag;//tile B
  wmma::fragment<wmma:accumulator,8,8,128, int> c_frag;  wmma::fill_fragment(c_frag,0);//tile C
  for (int i=0; i<(A_width/128);i++) {
    load_matrix_sync(a_frag, A+bx*8*A_width/32+i*128/32, A_width);//fetch tile A
    load_matrix_sync(b_frag, B+by*8*A_width/32+i*128/32, A_width);//fetch tile B
    bmma_sync(c_frag, a_frag, b_frag, c_frag); }//BMM
  for (int i=0; i<c_frag.num_elements;i++) c_frag.x[i]=A_width-2*c_frag.x[i];//fix for +1/-1 logic
  store_matrix_sync(C+(bx*8*B_width+by*8), c_frag, B_width, wmma::mem_row_major); //store tile C
}
BMM<<<dim3(A_height/16, B_width/8), dim3(32, 2)>>>(...); //invoke BMM kernel
\end{lstlisting}
\end{figure}

\vspace{2pt}\noindent\textbf{Design-1} Now we present our three BMM designs based on WMMA, which is the only API for operating the tensorcores. The baseline design is shown in Listing~\ref{lst:bmm1}. Each thread block comprises two warps and each warp processes BMM for a 128$\times$8 bit-tile in Line~\red{10}. Having two warps per thread block is for achieving the full occupancy of Turing SMs \cite{li2018warpconsolidation}. 

\begin{figure}[!t]
\begin{lstlisting}[caption=Bit-Matrix-Multiplication, captionpos='b', label={lst:bmm2}]
__global__ void BMM(unsigned *A, unsigned *B, int *C, int A_height, int A_width, int B_width) {
  using namespace nvcuda::wmma::experimental;
  __shared__ uint4 As[32], Bs[32];//buffering (8*128)*8 bit block in shared memory
  const int laneid=threadIdx.x;  const int wx=threadIdx.y;  const int wy=threadIdx.z; //tile index
  const int bx=blockIdx.x;  const int by=blockIdx.y; //block index
  wmma::fragment<wmma::matrix_a,8,8,128, precision::b1,wmma::row_major>a_frag;//tile A
  wmma::fragment<wmma::matrix_b,8,8,128, precision::b1,wmma::col_major>b_frag;//tile B
  wmma::fragment<wmma:accumulator,8,8,128, int> c_frag;  wmma::fill_fragment(c_frag,0);//tile C
  for(int k=0;k<A_width;k++){
    if(wx==0&&wy==0){//one warp fetches data into shared memory for 16 warps of a thread block
      As[laneid]=((uint4*)A)[(bx*32+laneid)*A_width+k]; Bs[laneid]=((uint4*)B)[(by*32+laneid)*A_width+k];}
    __syncthreads();//for respecting RAW dependency
    load_matrix_sync(a_frag, &As[wx*8], 128);  load_matrix_sync(b_frag, &Bs[wy*8], 128);
    bmma_sync(c_frag, a_frag, b_frag, c_frag);
    __syncthreads();}//for respecting WAR dependency
  for(int i=0; i<c_frag.num_elements; i++) c_frag.x[i] = (A_width*128)-(2*c_frag.x[i]);//+1/-1 BMM 
  store_matrix_sync(&C[(bx*4+wx)*8*B_width+(by*4+wy)*8], c_frag, B_width, wmma::mem_row_major);
}
BMM<<<dim3(A_height/32,B_width/32),dim3(32,4,4)>>>(...);
\end{lstlisting}
\end{figure}
\vspace{2pt}\noindent\textbf{Design-2} As memory load is the most important factor for matrix multiplication on GPU \cite{tan2011fast}, Design-2 aims at improving the efficiency of memory load. On one hand, using a whole warp to fetch only 128 bits is too lightweight. On the other hand, if coalescing memory access is enforced with each lane fetches 32bits, the total bit length becomes 32$\times$4$\times$8=1024 bits, which is probably too coarse-grained for a BNN FC layer given the matrix size is usually less than 2048. Therefore, motivated by \cite{jia2018dissecting}, we increase the load granularity per warp-lane to its max value of 128 bits, leveraging the effective \texttt{LDG.E.128} SASS instruction. With each lane fetching 128 bits, a warp of 32 lanes would fetch a bit segment of 4096 bits, which is sufficient for 4 warps to perform WMMA simultaneously. As a result, we use a representative warp to fetch 4096 bits of A and 4096 bits of B from global memory to shared memory, which are then dispatched to 16 warps for WMMA execution, as listed in Listing~\ref{lst:bmm2}. Essentially, each thread block processes a BMM of (128,32)$\times$(32,128) while each warp processes (128,8)$\times$(8,128). Line~\red{11}-\red{12} show how to invoke 128-bit global memory load through vectorization \cite{jia2018dissecting}. Note that \texttt{load\_matrix\_sync} is 5$\times$ faster on shared memory than global memory (Section~4.1).

%The second approach is proposed with a different angle. Typically, the matrix size in an FC layer is between 256 and 2048. Even for a 2048$\times$2048 matrix, with each warp processes 1024$\times$8 bit-tiles in Design-1, the entire workload is just enough for 512 warps. Given 32 warps per SM for Turing GPUs and 68 SMs in RTX2080Ti, the overall parallelism offered by the GPU hardware is 2176. In other words, 76\% of the SM warp-slots are idle. To tackle this low-utilization problem, in Design-2, we assign each warp with finer-grained workload at the cost of less data reuse. Essentially, each warp processes a 128

\begin{figure}[!t]
\begin{lstlisting}[caption=BMM in new format with binarized output, captionpos='b', label={lst:bmm3}]
#define FLIPBITS(n,b) ((n)^((1u<<(b))-1)) //flip the last b bits of n
__global__ void BMMb(unsigned *A, unsigned *B, unsigned *C, int A_height, int A_width, int B_width) {
  using namespace nvcuda::wmma::experimental;
  int bx = blockIdx.x*blockDim.y+threadIdx.y;  int by = blockIdx.y;
  __shared__ int Cs[64];
  unsigned laneid; asm("mov.u32 %0, %%laneid;":"=r"(laneid)); //get laneid
  wmma::fragment<wmma::matrix_a,8,8,128, precision::b1,wmma::row_major>a_frag;//tile A
  wmma::fragment<wmma::matrix_b,8,8,128, precision::b1,wmma::col_major>b_frag;//tile B
  wmma::fragment<wmma:accumulator,8,8,128, int> c_frag;  wmma::fill_fragment(c_frag,0);//tile C
  for (int i=0; i<(A_width/128);i++) {
    load_matrix_sync(a_frag, A+bx*8*A_width/32+i*128/32, A_width);//fetch tile A
    load_matrix_sync(b_frag, B+by*8*A_width/32+i*128/32, A_width);//fetch tile B
    bmma_sync(c_frag, a_frag, b_frag, c_frag); }//BMM
  for (int i=0; i<c_frag.num_elements;i++) c_frag.x[i]=A_width-2*c_frag.x[i];//fix for +1/-1 logic
  store_matrix_sync(Cs, c_frag, B_width, wmma::mem_row_major); } //store tile C
  union{unsigned data; uchar elements[4];} p0,p1;
  p0.data=__ballot(A_width-2*Cs[laneid]>=0);
  p1.data=__ballot(A_width-2*Cs[laneid+32]>=0); 
  if (laneid<4) { uchar* Cb = (uchar*)&C[0];
    Cb[(bx*8+laneid)*(B_width/8)+FLIPBITS(by,2)]=p0.elements[3-laneid];
    Cb[(bx*8+4+laneid)*(B_width/8)+FLIPBITS(by,2)]=p1.elements[3-laneid]; }
}
BMMb<<<dim3(A_height/16, B_width/8), dim3(32, 2)>>>(...); //invoke BMMb kernel
\end{lstlisting}
\end{figure}

\vspace{2pt}\noindent\textbf{Design-3} We adopt our new FSB format for BMM. Listing~\ref{lst:bmm3} shows the design with the output matrix binarized. As can be seen, after BMM, each warp holds a tile of C in size of 8$\times$8. We use \texttt{\_\_ballot()} to binarize the 64 elements using the entire 32 lanes in Line~\red{17-18}. In order to write an 8-bit \texttt{uchar} to a 32-bit \texttt{unsigned}, we define a \texttt{union} for packing and unpacking. As NVIDIA GPUs adopt little endian, we define \texttt{FLIPBIT()} for efficient byte-index translation.

\begin{figure}[!t]
\begin{lstlisting}[caption=BTC-based BConv design, captionpos='b', label={lst:bconv1}]
__global__ void BTC_Conv2d(const unsigned* __restrict__ input, 
        const unsigned* __restrict__ filter, int* output, const int in_channels, const int out_channels, 
        const int in_width, const int in_height, const int filter_width, const int filter_height, 
        const int batch, const int stride_vertical, const int stride_horizontal, const int out_width, 
        const int out_height, const int pad_h, const int pad_w) {
  using namespace nvcuda::wmma::experimental;
  wmma::fragment<wmma::matrix_a, 8, 8, 128, precision::b1, wmma::row_major> a_frag;
  wmma::fragment<wmma::matrix_b, 8, 8, 128, precision::b1, wmma::col_major> b_frag;
  wmma::fragment<wmma::accumulator, 8, 8, 128, int> c_frag;
  unsigned laneid; asm("mov.u32 %0, %%laneid;":"=r"(laneid)); //get laneid
  const int bx = blockIdx.x; //over Q: out_width 
  const int by = blockIdx.y; //over P: out_height
  const int bz = blockIdx.z; //over O: out_channel/8 * batch/8
  const int ins = (in_channels>>7);//number of steps in C: in_channels
  const int bn = bz/(out_channels>>3); //batch
  const int bo = bz%(out_channels>>3); //out_channel
  //coord (ax,ay) in Input from bx,by in Output
  const int ax0 = bx*stride_horizontal-pad_w; const int ay0 = by*stride_vertical-pad_h;
  int exclude = 0;  wmma::fill_fragment(c_frag, 0); //track the number of filter entries that are masked off
  for (int r=0; r<filter_height; r++) {  //load a window of data from Input
    for (int s=0; s<filter_width; s++) {
      const int ay = ay0 + r; //y-coord in Input
      const int ax = ax0 + s; //x-coord in Input
      if ((ay>=0) && (ay<in_height) && (ax>=0) && (ax<in_width)) { //within Input frame
        for (int c=0; c<ins; c++) { //per 128-bit in C:in_channels
         //input:[H,W,N,C], filter:[K,K,O,C]
         load_matrix_sync(a_frag,&input[(ay*in_width+ax)*batch*ins*4+bn*ins*32+c*4],in_channels);                                              
         load_matrix_sync(b_frag,&filter[(r*filter_width+s)*out_channels*ins*4+bo*ins*32+c*4],in_channels);
         bmma_sync(c_frag, a_frag, b_frag, c_frag);}}
      else exclude++; //accumulate for points not in frame
  } }
  for (int i=0; i<c_frag.num_elements; i++)
    c_frag.x[i] = in_channels*filter_height*filter_width - exclude*in_channels - (2*c_frag.x[i]);
  //output: [P,Q,N,O] => [by,bx,bn,bo]
  store_matrix_sync(&output[(by*out_width+bx)*batch*out_channels + bn*8*out_channels + bo*8],
          c_frag, out_channels, wmma::mem_row_major); }
\end{lstlisting}
\end{figure}

\subsection{BConv for Convolution Layer}
The convolution operation here is to cross-correlate a 4D input tensor (\emph{batch, input\_height, input\_width, input\_channels}) with a 4D weight tensor (\emph{weight\_height, weight\_width, input\_channels, output\_channels}). We use H to denote input\_height, W to denote input\_width, N to denote batch, C to denote input\_channel, K to denote weight size and O to denote output\_channel. \emph{TensorFlow} thus uses \texttt{NHWC} for input and \texttt{KKCO} for filter. \emph{PyTorch} uses \texttt{NCHW} for input and \texttt{OCKK} for filter. Traditionally, a 2D convolution can be transformed into GEMM through the \emph{im2col()} process \cite{chetlur2014cudnn, chellapilla2006high}, which can then be accelerated by the tensorcores. However, for BConv, directly converting to BMM is not feasible due to the challenge in padding \cite{electronics8060661}. Different from normal convolution where the padded zeros shall not affect the correctness, in BConv an element zero actually denotes -1. Therefore, after the \emph{im2col()} process, we are unable to distinguish the padded 0s from the meaningful zeros representing -1, leading to inaccurate convolution results. 

\begin{figure*}[!t]
\centering
\vspace{-5pt}
\includegraphics[width=1.8\columnwidth]{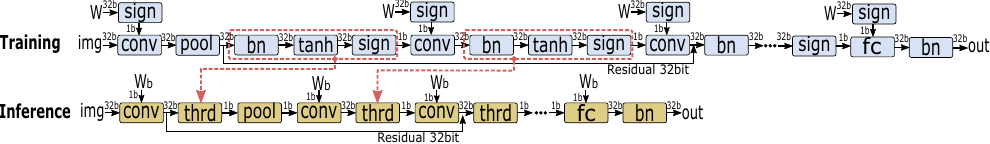} 
\caption{A typical network structure of a ResNet BNN including 3 Conv layers and 1 FC layer for training and inference.}
\label{fig:layer}
\end{figure*}

Thus, the objective here is how to design BConv so that the padding issue can be well-managed but at the same time can still be accelerated by the bit tensorcores of Turing GPUs. On one hand, motivated by existing work \cite{li2019bstc}, if the entire filter window is processed sequentially by a single GPU thread, a status variable can be allocated to track how many entries of the filter window fall out of the frame of the input image, which can be used later to make an amendment accordingly for ensuring the correctness of bit-padding. On the other hand, if we ignore the image size and filter size for now but looking at a particular point \emph{[i,j]} of the input image, the batch of N images at that point cross-correlating with an entry of the filter window \emph{[r,s]} is essentially to calculate the following output point \emph{[p,q]}:
\begin{equation}
\label{eq:conv}
Output_{[p,q]}=\sum_{k=1}^{C}input(N,k)_{[i,j]}\times filter(k,O)_{[r,s]}
\end{equation}
This is just equivalent to multiplying a bit matrix in size \emph{(N,C)} with another matrix in size \emph{(C,O)}, which can be performed by the bit-tensorcores. Consequently, our idea is to change the input tensor to \texttt{HWNC}, the filter tensor to \texttt{KKCO}, and perform BMM along their last two dimensions. 

Our first design is shown in Listing~\ref{lst:bconv1}. We use each warp to traverse the input\_channel space at Line~\red{28} and perform the computation for 8 input\_images over 8 output\_channels (i.e., \emph{(8,C)}$\times$\emph{(C,8)}) using the bit-tensorcores in Line~\red{30-32}. We use "\emph{exclude}" to track the number of entries outside the filter frame at Line~\red{33} and amendment the results at Line~\red{36} for padding and the $\pm$1 logic (see Eq~\ref{eq:dot-product}). We use \emph{c\_frag} for storing the partial results of convolution. Eventually, the 8$\times$8 resulting matrix tile stored in \emph{c\_frag} is written back to the global memory in row-major at Line~\red{38-39}.

Our second BConv design leverages the new bit data format. We reform the last two dimensions of the input tensor \emph{(N,C)} in a bit-tile of 128$\times$8 bits in row-major, and the filter tensor \emph{(C,O)} in a bit-tile of 128$\times$8 bits in column-major. Then, we can adjust the \emph{ldm} in Line~\red{30-31} from "\emph{in\_channels}" to 128, later we will show the impact of this adjustment.

\section{BNN Design with BTC}

%We present BNN structure and our BTC-based BNN design.

%We present the overall BNN structure and our BTC-based BNN implementation.

% \blue{We also discuss how to form the BTC-based BNNs into a \emph{Binary Ensemble Neural Network} (BENN).}

\subsection{BNN Network Structure}

Figure~\ref{fig:layer} illustrates the network structure of an example ResNet. To avoid losing too much non-recoverable information at the beginning, if the input images are in full-precision (e.g., after preprocessing), the first layer of BNN is not binarized \cite{rastegari2016xnor, hubara2016binarized, tang2017train}. BWN is adopted here in which only the weight matrix is binarized. Consequently, we are unable to use BTC to accelerate the first layer. Also because the input channels of the first layer is usually very small (e.g., red, green, blue), to avoid alignment issue and fully leverage data locality, we binarized the weight matrix into a 4D bit tensor in \texttt{KKCO} format and buffer the weight into the shared memory for reuse. Then, by extracting each bit of the weight, depending on whether it is 1 or 0, we add or subtract the corresponding element of the input matrix. The output matrix is binarized and stored in particular bit-format as the input for the next layer.

% \red{In an FC layer, existing work has already shown that batch normalization, which is crucial for BNNs, can be integrated with the binarization (i.e., Eq~\ref{eq:binarize}) of the next layer to constitute a threshold-comparison operation (i.e., returns 1 if greater than a threshold $\tau$ and 0 otherwise) \cite{umuroglu2017finn, li2019bstc}. This threshold comparison can be further combined with the BMM here so that the memory store of the current layer and the memory load of the next layer can be performed on binary data rather than full-precision data, significantly reducing memory access. }

Shown in Figure~\ref{fig:layer}, regarding training, a BNN convolution layer typically comprises \emph{binarization} (\texttt{sign}), \emph{bit-convolution} (\texttt{conv}), \emph{batch-normalization} (\texttt{bn}), \emph{hard-tanh} (\texttt{tanh}), and \emph{pooling} (\texttt{pool}). Binarization is the \texttt{sign} function following Eq~\ref{eq:binarize}. Batch-normalization \cite{ioffe2015batch} is to reduce the batch noise:
\begin{equation}
 y_{i,j} = \left( \frac{x_{i,j}-\mathbb{E}[x_{*,j}]}{\sqrt{Var[x_{*,j}]+\epsilon}} \right)\cdot \gamma_{j} + \beta_{j}
\label{eq:normalization}
\end{equation}
Note that \texttt{bn} is essential for BNNs, as missing it will render the training unable to converge. Additionally, having \texttt{bn} brings two extra benefits: (1) bias is thus not necessary for the bit convolution or fully-connected layer, as bias can be integrated with $\beta_{j}$ in Eq~\ref{eq:normalization}; (2) the scaling layer proposed in \cite{rastegari2016xnor, tang2017train} for BNN is also not necessary as it can be integrated with $\gamma_{j}$ in Eq~\ref{eq:normalization}. Hard \texttt{tanh} is a piecewise linear function:
\begin{equation}
Htanh(x) = Clip(x,-1,1)
\label{eq:tanh}
\end{equation}
Since \texttt{tanh} is immediately followed by the \texttt{sign} function, it has none effect on inference or the forward pass of training. The major purpose of \texttt{tanh} is to constrain the gradient of the \texttt{sign} function between -1 and +1 in the backward pass \cite{courbariaux2016binarized}. Otherwise, if the full-precision activation is too large, the gradient will be zeroed-out. Additionally, since the \texttt{sign} binarization function has already imposed non-linearity into the network, no other activation function such as \emph{ReLU} \cite{rastegari2016xnor} and \emph{PReLU} \cite{tang2017train} is actually needed for BNN. Conversely, extra activation functions might be harmful based on our tests. 

Regarding the order of these unit functions, it should be \texttt{tanh$\to$sign$\to$bconv$\to$pool$\to$bn$\to$tanh$\to$sign} for the training, as it has already been shown that placing \texttt{pool} before \texttt{bn} can lead to increased training accuracy \cite{rastegari2016xnor, electronics8060661}. However, for inference it would be much faster if equivalently \texttt{pool} is located after \texttt{bn} and even the binarization of the next layer to convert a max pooling into a logic-\emph{OR} operation \cite{umuroglu2017finn, li2019bstc}. Additionally, for inference, \texttt{bn} and \texttt{sign} of the next layer can be aggregated as a simple threshold comparison operation (i.e., returns +1 if greater than a threshold $\tau$ and -1 otherwise) \cite{umuroglu2017finn, li2019bstc}, labeled as \texttt{thrd} in Figure~\ref{fig:layer}. In this way, \emph{thrd} can be further fused with \emph{bconv} or \emph{bmm} to reduce the volume of data access if the residual is not saved. Finally, \texttt{tanh} is not required for inference as discussed. Consequently, the ultimate function order becomes \texttt{thrd$\to$bconv$\to$thrd$\to$pool$\to$bconv} for inference. Similar condition is also applied for the FC layers.

%It means if the result of \emph{Bconv} or \emph{BMM} is larger than a threshold, it outputs +1; otherwise it outputs -1.

Traditionally, BNN's last layer is also in full-precision \cite{courbariaux2016binarized, rastegari2016xnor}. However, Tang et al. \cite{tang2017train} showed that binarizing the final layer with a learned scaling layer could significantly compact the model as FC layers comprises the most parameters. Our observation here is that such a scaling layer can be absorbed by adding a \texttt{bn} function for the last layer, which may provide even better performance due to more constraint output range for the following \texttt{softmax} function. Note, for the last layer, since the output is real-valued and there is no future binarization, \emph{bn} cannot be converted into a \emph{thrd} function.

In terms of more advanced models such as ResNet, to avoid gradient diminishing or explosion, the cross-layer shortcut connections become vital. Here, the main performance concern is that these residuals are real-valued (bit-residual cannot convey gradient), which may incur substantial extra memory load \& store compared with directly saving the bits after \texttt{thrd}. In addition, the residual may need a pooling layer before the injection. Furthermore, it is also possible that the number of channels needs to adjust. In those scenarios, we use the type-A shortcut of ResNet \cite{he2016deep}. 

%Note that the need for conserving the real-valued residual is the reason in Section~V we discuss the designs where \emph{C} is not binarized.

\begin{table*}[!t]
\centering\footnotesize
%\vspace{-20pt}
\caption{Evaluation Platforms. "Reg" refers to the number of 4-byte registers. "Thds" refer to threads. "Dri/Rtm" refer to CUDA driver and runtime versions.}
\begin{tabular}{|c|c|c|c|c|c|c|c|c|c|c|c|c|}
\hline
\textbf{GPU} & \textbf{Arch/CC} & \textbf{Code}  & \textbf{SMs} & \textbf{CTAs/SM} & \textbf{Warps/SM} & \textbf{Thds/CTA} & \textbf{Regs/SM}  & \textbf{Shared/SM} & \textbf{TCUs/SM} & \textbf{Memory}  & \textbf{Mem Bandwidth} & \textbf{Dri/Rtm} \\ \hline
RTX-2080Ti & Turing-7.5 & TU102 &  68 & 16 & 32 & 1024 & 64K & 64K & 8 & 11GB GDDR6 & 616 GB/s  & 10.1/10.0  \\ \hline
RTX-2080 & Turing-7.5 & TU104 & 46 & 16 & 32 & 1024 & 64K & 64K & 8 & 8GB GDDR6 & 448 GB/s  & 10.0/10.0  \\ \hline
\end{tabular}
\label{tab:gpu}
\end{table*}

\subsection{BTC based Implementation}

Similar to \cite{li2019bstc}, we have also fused all the layer functions into a single GPU kernel so the repeated kernel invocation \& release overhead (as long as 20 $\mu s$ per invocation \cite{lebeane2017gpu}) can be eliminated. We implement each layer function as a GPU device function. These device functions are called from a global function where the BNN network model is defined. Due to data dependency across the layers, to ensure consistency, we rely on CUDA's cooperative-groups for global synchronization among all SMs. There are two major challenges for the overall design here: (i) \emph{Achieving high SM utilization}. Since \emph{WMMA} is executed at the warp level, with 32 warps per SM for Turing GPUs and 68 SMs in RTX2080Ti for instance, the overall parallelism offered by the hardware is 2176 warps, implying 2176 BMMs sized (8,128)$\times$(128,8) per round. Consequently, the task granularity per warp should be as small as possible \cite{li2018warpconsolidation} in order to use all the SM warp slots and achieve workload balance; (ii) \emph{Adapting to WMMA format}. As BTC can only process BMM sized (8,128)$\times$(128,8)=(8,8), we need to ensure the row of the FC input matrix, the column of the weight, the batch of the BConv image, and the output channel can all divide 8, while the column of the FC input matrix, the row of the weight, the input channel of BConv can all divide 128. Both requirements need to be consistent across all layers. Given the BNN model can be in arbitrary configuration and we internally use our own FSB format, the address translation and calculation become more complicated. There is another format change after the final Conv layer and ahead of the first FC layer to ensure correct format transition.

%The second approach is proposed with a different angle. Typically, the matrix size in an FC layer is between 256 and 2048. Even for a 2048$\times$2048 matrix, with each warp processes 1024$\times$8 bit-tiles in Design-1, the entire workload is just enough for 512 warps. Given 32 warps per SM for Turing GPUs and 68 SMs in RTX2080Ti, the overall parallelism offered by the GPU hardware is 2176. In other words, 76\% of the SM warp-slots are idle. To tackle this low-utilization problem, in Design-2, we assign each warp with finer-grained workload at the cost of less data reuse. Essentially, each warp processes a 128

%Ensemble methods combine multiple hypotheses, hoping to form a better one than the best hypothesis alone. It uses multiple weak learners to build a strong learner [58]. Boosting can be seen as a linear regression where data features are the input to the weak learner h, and the output of the boosting is the weighted summation of these h functions. Bagging (bootstrap aggregating) is a method to improve the generality of the predictive model to reduce over-fitting. It is based on a model-averaging technique. 

\section{Evaluation}

%We describe the experiment configurations, show the evaluation results, and discuss the performance of BNN using different models and datasets.

%We evaluate our BTC-based BNN design. We first describe the experiment configurations. Then, we show the evaluation results for BMM and BConv. Finally, we discuss the performance of BNN using different models and datasets.

\subsection{Experiment Configuration}
We use two NVIDIA Turing GPUs with CC-7.5 for evaluation. Their information is listed in Table~\ref{tab:gpu}. The RTX2080 GPU is in a Linux 3.10.0 system with Intel Xeon E5-2680 CPU @ 2.80 GHz, 128 GB DDR3 DRAM and gcc-4.8.5. The RTX2080Ti GPU is in a Linux 2.6.32 system with Intel Xeon E5-6230 CPU @ 2.10 GHz, 384 GB DDR4 and gcc-4.8.5. All the results reported are the average of 10 times' execution.

%More detailed information can be found in the appendix.

\subsection{BMM Evaluation}

\begin{table}[!t]
\centering\scriptsize
\caption{BMM full-precision output schemes.}
\begin{tabular}{|c|l|c|c|c|}
\hline
 \textbf{Schemes} & \textbf{Description} & \textbf{Algorithm} & \textbf{Input}  & \textbf{Output}  \\ \hline
 \emph{cuBLAS} & Simulating BMM via FP16 HGEMM & HGEMM & 16bit & 32bit \\ \hline
 \emph{xnor} & BMM design in \cite{courbariaux2016binarized} & BMM & 32bit & 32bit \\ \hline
 \emph{bmm32} & 32bit BSTC BMM in \cite{li2019bstc} & BMM & 32bit & 32bit \\ \hline
 \emph{bmm64} & 64bit BSTC BMM in \cite{li2019bstc} & BMM & 32bit & 32bit \\ \hline
 \emph{bmms32} & Fine-grained 32bit BSTC BMM in \cite{li2019bstc} & BMM & 32bit & 32bit \\ \hline
 \emph{bmms64} & Fine-grained 64bit BSTC BMM in \cite{li2019bstc} & BMM & 32bit & 32bit \\ \hline
 \emph{cutlass} & BMM on TCUs in \emph{Cutlass} library \cite{cutlass} & BMM & 1bit & 32bit \\ \hline
\emph{u4} & BMM via uint-4 MM on TCUs in \emph{Cutlass} \cite{cutlass} & 4bit-MM & 4bit & 32bit \\ \hline
 \emph{bmma} & Design-1: basic BTC implementation & BMM & 32bit & 32bit \\ \hline
 \emph{bmma128} & Design-2: 128bit load and shared memory & BMM  & 32bit & 32bit     \\ \hline
 \emph{bmmafmt} & Design-3: new format & BMM & 32bit & 32bit \\ \hline
\end{tabular}
\label{tab:BMM_schemes}
\end{table}

\begin{table}[!t]
\centering\scriptsize
\caption{BMM bit output schemes.}
\begin{tabular}{|c|l|c|c|}
\hline
 \textbf{Schemes} & \textbf{Description} & \textbf{Input}  & \textbf{Output}  \\ \hline
 \emph{bmm32\_b} & 32bit BSTC BMM in \cite{li2019bstc} with bin output  & 1bit & 1bit \\ \hline
 \emph{bmm64\_b} & 32bit BSTC BMM in \cite{li2019bstc} with bin output  & 1bit & 1bit \\ \hline
 \emph{bmms32\_b} & Fine-grained 32bit BSTC BMM in \cite{li2019bstc} with bin output &  1bit & 1bit \\ \hline
 \emph{bmms64\_b} & Fine-grained 64bit BSTC BMM in \cite{li2019bstc} with bin output &  1bit & 1bit \\ \hline
 \emph{bmma\_b} & Design-1: basic BTC implementation with bin output &  1bit & 1bit \\ \hline
 \emph{bmma128\_b} & Design-2: 128bit load with bin output  & 1bit & 1bit     \\ \hline
 \emph{bmmafmt\_b} & Design-3: new format with bin output & 1bit & 1bit \\ \hline
 \end{tabular}
\label{tab:BMMb_schemes}
\end{table}

\begin{figure*}[!htb]
\minipage{0.5\columnwidth}
\includegraphics[width=1.05\columnwidth]{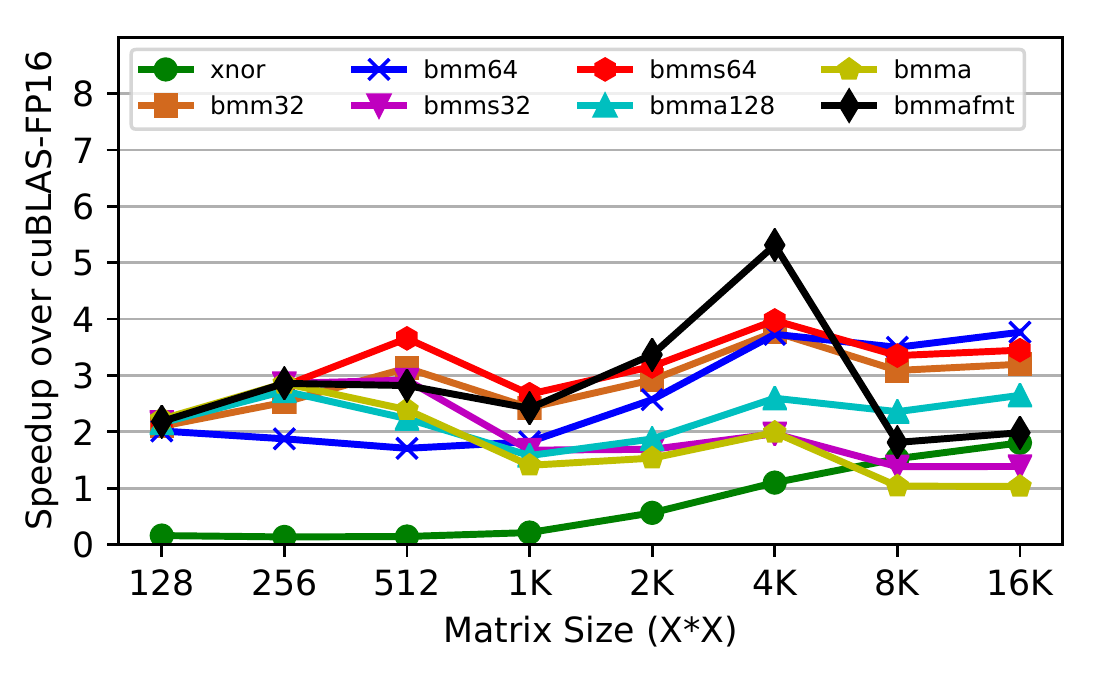} 
\caption{\blue{General BMM on RTX2080}}
\label{fig:bmm_rtx2080}
\endminipage\hfill
\minipage{0.5\columnwidth}
\includegraphics[width=1.05\columnwidth]{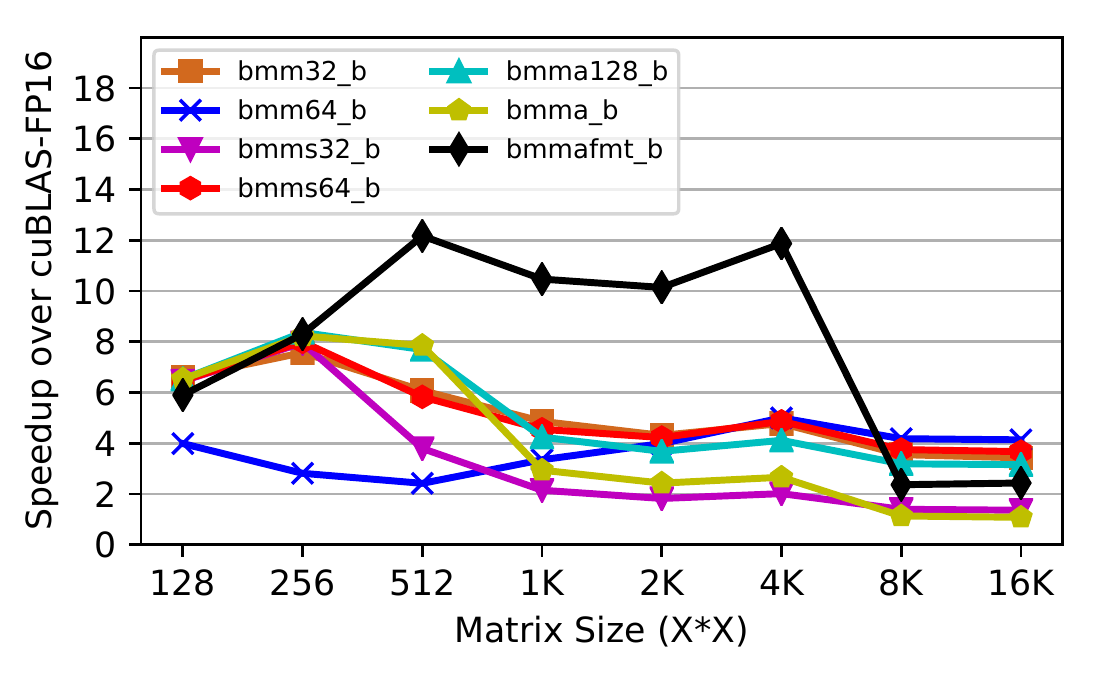} 
\caption{\blue{Specific BMM on RTX2080}}
\label{fig:bmmb_rtx2080}
\endminipage\hfill
\minipage{0.5\columnwidth}
\includegraphics[width=1.05\columnwidth]{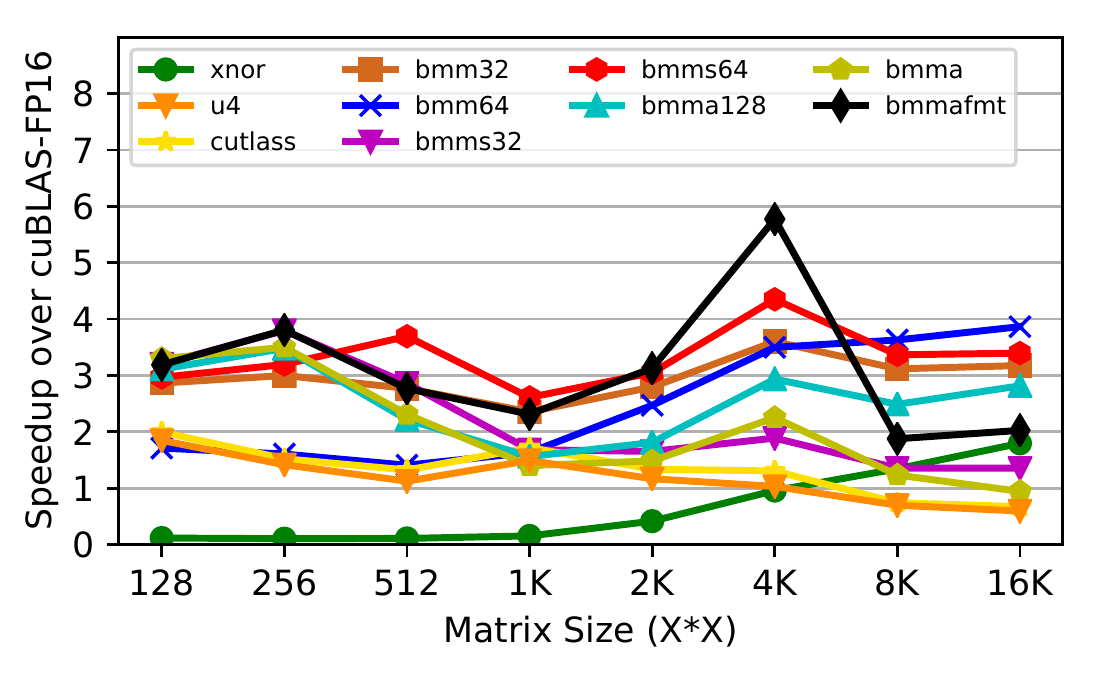} 
\caption{\blue{General BMM on RTX2080Ti}}
\label{fig:bmm_rtx2080ti}
\endminipage\hfill
\minipage{0.5\columnwidth}
\includegraphics[width=1.05\columnwidth]{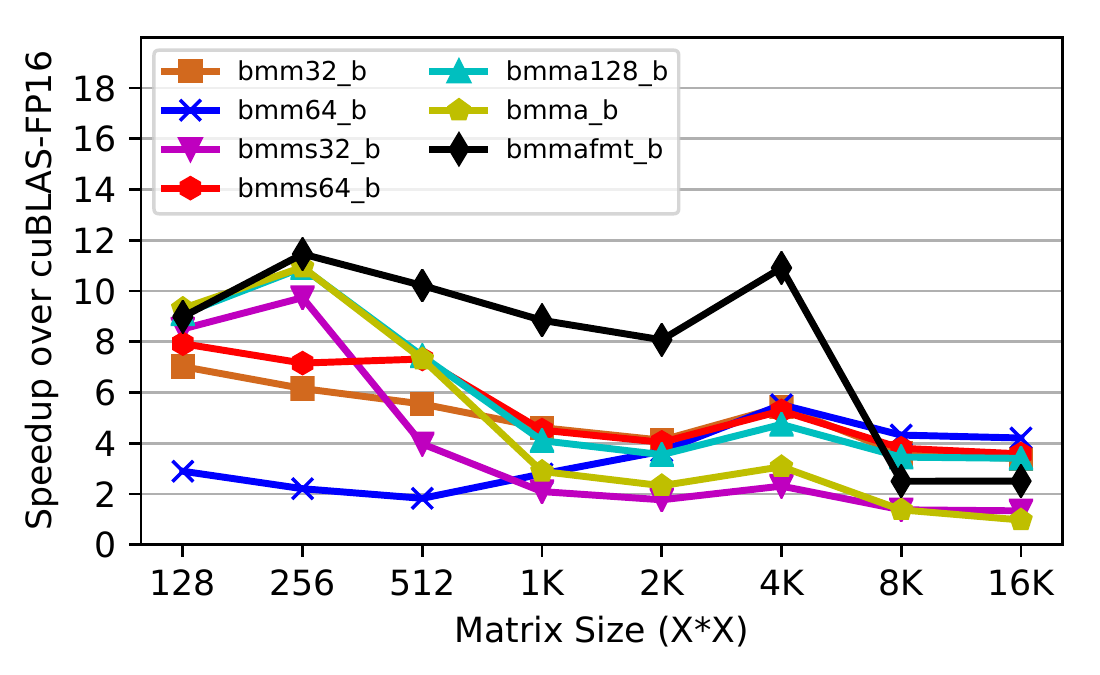} 
\caption{\blue{Specific BMM on RTX2080Ti}}
\label{fig:bmmb_rtx2080ti}
\endminipage
\end{figure*}

%\begin{figure*}[!htb]
%\minipage{0.5\columnwidth}
%\includegraphics[width=1.05\columnwidth]{figure/bmm_rtx2080_oneside.pdf} 
%\caption{\blue{General non-square BMM on RTX2080}}
%\label{fig:bmm_rtx2080}
%\endminipage\hfill
%\minipage{0.5\columnwidth}
%\includegraphics[width=1.05\columnwidth]{figure/bmm_rtx2080_bin_oneside.pdf} 
%\caption{\blue{Specific non-square BMM on RTX2080}}
%\label{fig:bmmb_rtx2080}
%\endminipage\hfill
%\minipage{0.5\columnwidth}
%\includegraphics[width=1.05\columnwidth]{figure/bmm_rtx2080ti_oneside.pdf} 
%\caption{\blue{General non-square BMM on RTX2080Ti}}
%\label{fig:bmm_rtx2080ti}
%\endminipage\hfill
%\minipage{0.5\columnwidth}
%\includegraphics[width=1.05\columnwidth]{figure/bmm_rtx2080ti_bin_oneside.pdf} 
%\caption{\blue{Specific non-square BMM on RTX2080Ti}}
%\label{fig:bmmb_rtx2080ti}
%\endminipage
%\end{figure*}

For BMM evaluation, we randomly generate square matrices with increased sizes from from 128 until 16K. \blue{We use the well-optimized 16-bit HGEMM (accelerated by TCUs) from \emph{cuBLAS} as the baseline.} We compare our three BTC-based BMM designs with the BMM approach from \cite{rastegari2016xnor}, the four BSTC BMM designs from \cite{li2019bstc}, and the BTC uint-4 and BMM designs from \emph{Cutlass} \cite{cutlass}. Note that both the \emph{Cutlass} BMM and uint-4 designs are accelerated by TCUs. They are the only available reference designs on BMM and uint-4 MM upon TCUs when performing the evaluation. We conduct two types of testing: (1) \textbf{General BMM} where both the input matrices and the output matrix are floating-points. It includes binarization for A and B, but excludes the binarization for C. The tested schemes are listed in Table~\ref{tab:BMM_schemes}. (2) \textbf{BNN-specific BMM} where both the input and the output matrices are binarized. It thus includes binarization for C but excludes A and B. This test reflects how BMM behaves in a BNN FC layer. The schemes are listed in Table~\ref{tab:BMMb_schemes}.

\begin{figure*}[!htb]
\minipage{0.5\columnwidth}
\includegraphics[width=1.05\columnwidth]{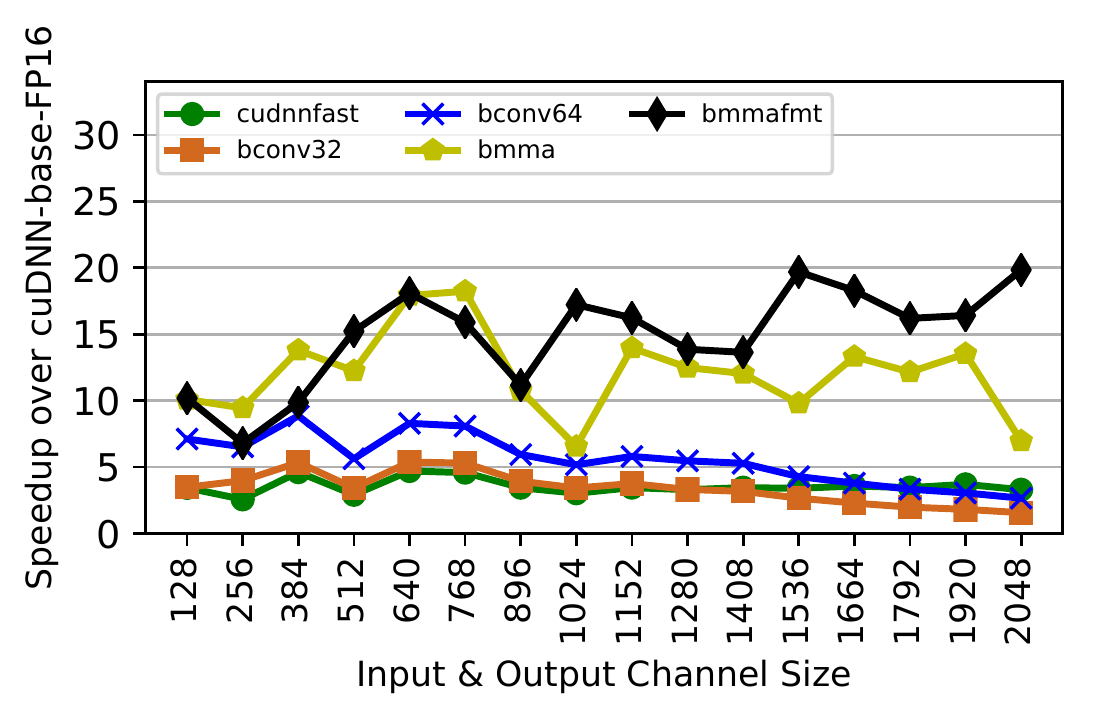} 
\caption{\blue{General BConv on RTX2080}}
\label{fig:bconv_rtx2080}
\endminipage\hfill
\minipage{0.5\columnwidth}
\includegraphics[width=1.05\columnwidth]{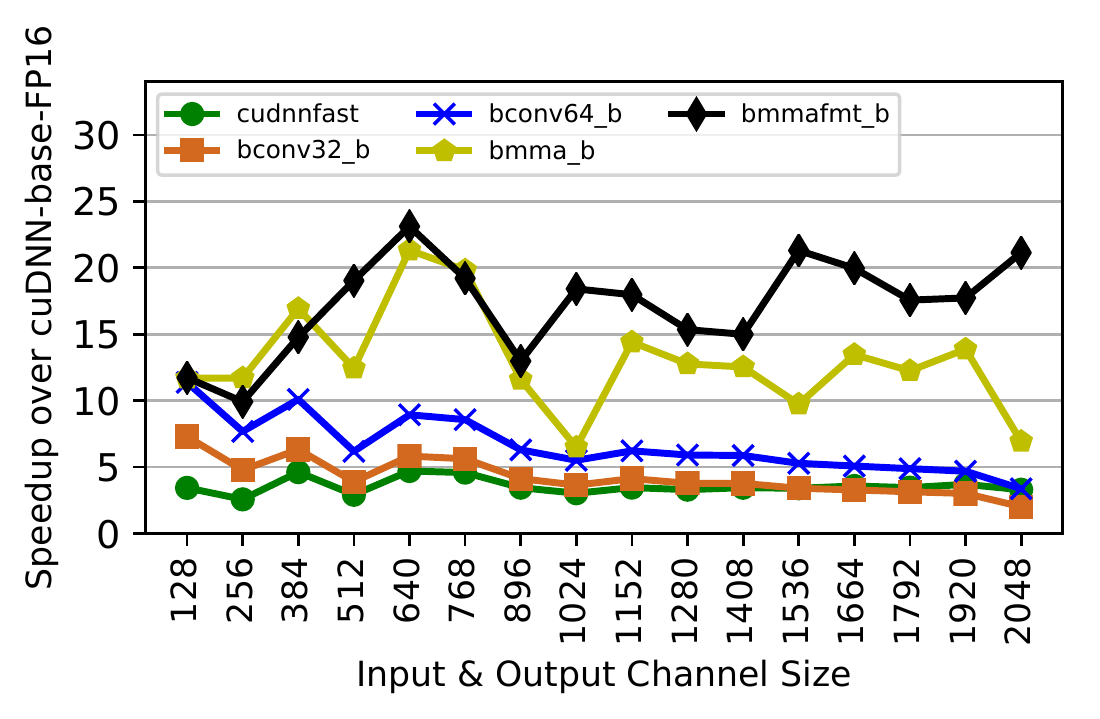} 
\caption{\blue{Specific BConv on RTX2080}}
\label{fig:bconvb_rtx2080}
\endminipage\hfill
\minipage{0.5\columnwidth}
\includegraphics[width=1.05\columnwidth]{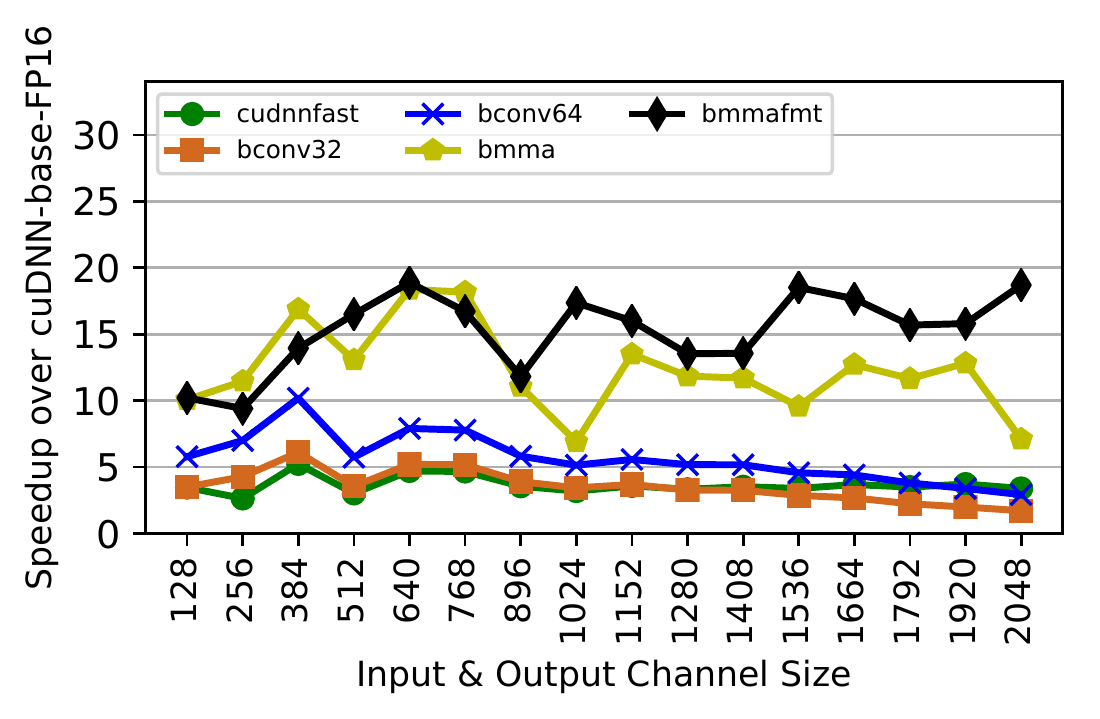} 
\caption{\blue{General BConv on RTX2080Ti}}
\label{fig:bconv_rtx2080ti}
\endminipage\hfill
\minipage{0.5\columnwidth}
\includegraphics[width=1.05\columnwidth]{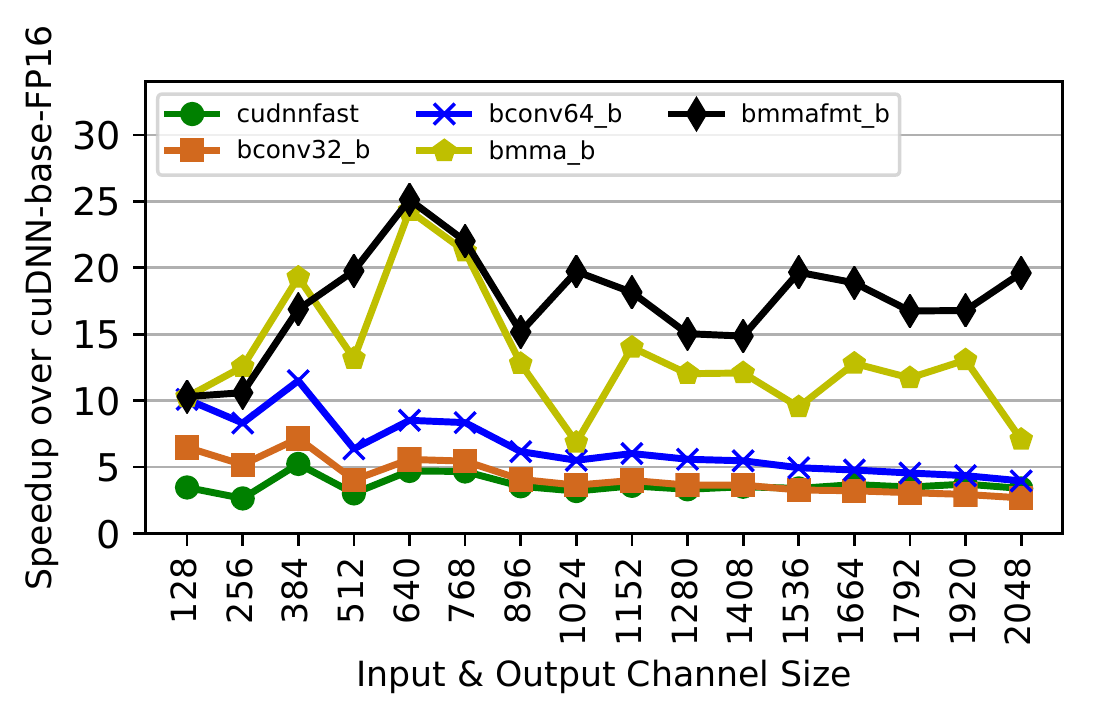} 
\caption{\blue{Specific BConv on RTX2080Ti}}
\label{fig:bconvb_rtx2080ti}
\endminipage
\end{figure*}

Figure~\ref{fig:bmm_rtx2080}, \ref{fig:bmmb_rtx2080}, \ref{fig:bmm_rtx2080ti} and \ref{fig:bmmb_rtx2080ti} show the results of the two BMM tests on TU104 RTX2080 GPU and TU102 RTX2080Ti GPU, respectively. For general BMM in Figure~\ref{fig:bmm_rtx2080} and \ref{fig:bmm_rtx2080ti}, we have three major observations: (I) No single approach dominates the entire matrix range --- For small matrices (n$\le$1K), the fine-grained 64bit BSTC is relatively better although the advantage is marginal. This might be due to more fine-grained thread-block tasks to leverage all SMs; For medium matrices (1K$<$n$\le$4K), Design-3 based on the proposed FSB-format obtains the best performance, particularly at 4K. For large matrices (n$>$4K), the performance of all BTC based designs drop. This is due to the fierce competition in BTC and reduced data reuse in the L0/L1 cache \cite{li2015adaptive}. Nevertheless, the size of FC layers of most BMMs fall in the medium range.
(II) Comparing among Design-1, 2, and 3, while Design-2 is always better than Design-1 due to improved load efficiency and shared memory reuse, the new-format based Design-3 significantly outperforms Design-1/2 except on very large matrices. Overall, without this new FSB format, BTC may not deliver any performance advantage over existing BSTC software solutions. For BNN-specific BMM in Figure~\ref{fig:bmmb_rtx2080} and \ref{fig:bmmb_rtx2080ti}, the avoidance of binarizing A \& B, and reduced memory store after binarizing C, dramatically amplify the supremacy of Design-3. \blue{The speedup is more than 12$\times$ over the FP-16 cuBLAS at 4K on RTX2080.}
 (III) Comparing between BMMs and uint-4 based GEMM over the same TCUs, we can observe that BMMs demonstrate obvious advantage. This is largely because (a) the smaller memory footprint (1-bit vs 4-bits) reduces the bandwidth and storage pressure over the data-path \cite{li2015transit, li2016x} and registers \cite{li2016critical}; (b) with the same bit-width for the ALUs in the TCUs, using 1-bit can compact 4$\times$ more data elements than using int-4 or uint-4. Similar conditions apply to other types such as int-8 and FP16.

\subsection{BConv Evaluation}

\begin{table*}[!t]
\centering\scriptsize
\caption{BTC Evaluation. ``1024FC'' refers to a fully-connected layer with 1024 neutrons. ``2x128C3'' refers to 2 convolution layer with 128 output channels and 3x3 filter. ``P2'' refers to a 2x2 pooling layer. ``128C11/4'' refers to a convolution layer with 128 output channels, 11x11 filter size and stride=4. "Input size" is of \emph{input\_height}$\times$\emph{input\_width}$\times$\emph{input\_channels} format. "Output" is the number of categories to classify. ``Ref'' is short for references. ``BNN'' refers to state-of-the-art BNN training accuracy from existing work. ``Our BNN'' is the BNN training accuracy we obtained from our own BNN implementation. ``Full-Precision'' is the 32 bits full-precision training accuracy from existing works.}
\begin{tabular}{|c|c|c|c|l|c|c|c|c|c|c|}
\hline
\textbf{Dataset} & \textbf{Ref} & \textbf{Network} & \textbf{Ref} & \textbf{Network Structure} & \textbf{Input Size} & \textbf{Out} & \textbf{BNN} & \textbf{Our BNN} & \textbf{Full-Precision}   \\ \hline
\emph{MNIST} & \cite{lecun2010mnist} & MLP & \cite{courbariaux2016binarized}  & \textsf{1024FC-1024FC-1024FC-1024FC} &  28x28x1 & 10 & 98.6\% \cite{courbariaux2016binarized} & 97.6\% & 99.1\%\cite{courbariaux2016binarized} \\ \hline
\emph{Cifar-10} & \cite{krizhevsky2014cifar} & VGG & \cite{courbariaux2015binaryconnect} & \textsf{(2x128C3)-MP2-(2x256C3)-MP2-(2x512C3)-MP2-(3x1024FC)} &  32x32x3 & 10  & 89.9\% \cite{courbariaux2016binarized} & 88.7\% & 90.9\%\cite{darabi2018bnn} \\ \hline
\emph{Cifar-10} & \cite{krizhevsky2014cifar} & ResNet-14 & \cite{rastegari2016xnor}  & \textsf{128C3/2-4x128C3-4x256C3-4x512C3-(2x512FC)} &  32x32x3 & 10 & N/A  & 91.6\% & N/A  \\ \hline
\emph{ImageNet} & \cite{deng2009imagenet} & AlexNet & \cite{krizhevsky2012imagenet} & \textsf{(128C11/4)-P2-(256C5)-P2-(3x256C3)-P2-(3x4096FC)} & 224x224x3 & 1000 & 75.7/46.1\%\cite{darabi2018bnn} & 74.2/44.7\% & 80.2/56.6\% \cite{darabi2018bnn} \\ \hline
%========
\emph{ImageNet} & \cite{deng2009imagenet} & VGG-16 & \cite{simonyan2014very} & \textsf{(2x64C3)-P2-(2x128C3)-P2-(3x256C3)-P2-2x(3x512C3-P2)-(3x4096FC)} & 224x224x3 & 1000 & 76.8\%/NA \cite{hubitflow2018} & 77.7/53.4\% & 88.4\%/NA \cite{hubitflow2018} \\ \hline
%========
\emph{ImageNet} & \cite{deng2009imagenet} & ResNet-18 & \cite{rastegari2016xnor} & \textsf{64C7/4-4x64C3-4x128C3-4x256C3-4x512C3-(2x512FC)} & 224x224x3 & 1000 & 73.2/51.2\% \cite{rastegari2016xnor} & 72.7/48.6\%  &  89.2/69.3\% \cite{rastegari2016xnor} \\ \hline
\end{tabular}
\label{tab:bhtc_evaluation}
\end{table*}

For BConv, there are much more parameters than BMM: \emph{input\_height}, \emph{input\_width}, \emph{weight\_height}, \emph{weight\_width}, \emph{batch}, \emph{input\_channels}, \emph{output\_channels}, \emph{stride}, \emph{pooling}, etc. We compare our two BTC-based designs (Note that we use \emph{bmma} to denote Design-1 and \emph{bmmafmt} to denote Design-2) \blue{with TCU-accelerated half-precision} \emph{cuDNN-base} (no workspace), \emph{cudnn-fast} (plenty workspace), and two BSTC (\emph{bconv32} and \emph{bconv64}) designs \cite{li2019bstc}. We use cuDNN-base as the baseline and perform the two types of test: (1) \textbf{General BConv} where the input, filter and output tensors are all floating-points; (2) \textbf{BNN-specific BConv} where all of them are binarized. 

%The schemes are listed in Table~\ref{tab:BConvb_schemes}.

%The schemes are listed in Table~\ref{tab:BConv_schemes}

Figure~\ref{fig:bconv_rtx2080}, \ref{fig:bconvb_rtx2080}, \ref{fig:bconv_rtx2080ti} and \ref{fig:bconvb_rtx2080ti} show the results of the two types of tests with \emph{batch}=16, \emph{input\_size}=64, \emph{weight\_height}=3 and \emph{stride}=1 on the two Turing GPUs. We increase both \emph{input\_channels} (C) and \emph{output\_channels} (O) from 128 to 2048. As is shown, our two BTC-based approaches exhibit considerable speedups over existing methods. Particularly, the FSB new format design achieves up to 25$\times$ over \blue{the FP-16 cuDNN} with C=O=640 on RTX2080Ti. Comparing between the two BTC designs, we can see that (i) when C=O=128, the two designs are just equivalent, so they show similar performance; (ii) When C=O=384, Design-1 is better possibly because \emph{ldm=384} is also a good choice for memory load (see Section~4). (iii) For the other points, Design-2 shows obvious advantages.

\subsection{BNN Evaluation}

\begin{table*}[!t]
\centering\scriptsize
\caption{BTC Inference Performance on NVIDIA Turing RTX2080 GPU.}
\begin{tabular}{|c|c|c|c|c|c|c|c|c|c|c|c|c|}
\hline
 & \multicolumn{2}{c|}{\bf MNIST-MLP} &  \multicolumn{2}{c|}{\bf Cifar10-VGG} &  \multicolumn{2}{c|}{\bf Cifar10-ResNet14} &  \multicolumn{2}{c|}{\bf ImageNet-AlexNet} &  \multicolumn{2}{c|}{\bf ImageNet-VGG} &  \multicolumn{2}{c|}{\bf ImageNet-ResNet18} \\\hline
 \textbf{Schemes} & \textbf{8 Latency} & \textbf{Throughput} & \textbf{8 Latency}  & \textbf{Throughput} & \textbf{8 Latency} & \textbf{Throughput} & \textbf{8 Latency}  & \textbf{Throughput} & \textbf{8 Latency} & \textbf{Throughput} & \textbf{8 Latency}  & \textbf{Throughput} \\ \hline
 
\textbf{SBNN-32} & 0.227ms & 2.88$\times10^6$fps & 1.891ms & 1.06$\times10^4$fps & 5.138ms & 4.17$\times10^3$fps & 4.494ms & 3.18$\times10^3$fps & 27.638ms & 4.26$\times10^2$fps & 6.550ms & 2.60$\times10^3$fps \\\hline
\textbf{SBNN-32-Fine} & 0.082ms & 1.97$\times10^6$fps & 1.536ms & 1.05$\times10^4$fps & 4.382ms & 3.97$\times10^3$fps & 3.928ms & 3.05$\times10^3$fps & 27.009ms & 4.25$\times10^2$fps & 5.944ms & 2.37$\times10^3$fps \\\hline
\textbf{SBNN-64} & 0.908ms & 8.44$\times10^4$fps & 2.816ms & 1.06$\times10^4$fps & 8.132ms & 4.59$\times10^3$fps & 17.258ms & 1.95$\times10^3$fps & 40.247ms & 5.23$\times10^2$fps & 8.108ms & 3.01$\times10^3$fps \\\hline
\textbf{SBNN-64-Fine} & 0.074ms & 5.51$\times10^6$fps & 0.999ms & 1.63$\times10^4$fps & 2.550ms & 6.52$\times10^3$fps & 2.871ms & 3.79$\times10^3$fps & 16.68ms & 6.65$\times10^2$fps & 3.736ms & 3.42$\times10^3$fps \\\hline
\textbf{BTC} & 0.061ms & 3.37$\times10^6$fps & 0.364ms & 3.62$\times10^4$fps & 0.827ms & 1.58$\times10^4$fps & 2.367ms & 3.85$\times10^3$fps & 7.449ms & 1.24$\times10^3$fps & 1.869ms & 5.48$\times10^3$fps \\\hline
\textbf{BTC-FMT} & 0.055ms & 5.48$\times10^6$fps & 0.338ms & 3.85$\times10^4$fps & 0.724ms & 1.71$\times10^4$fps & 2.326ms & 3.77$\times10^3$fps & 7.021ms & 1.34$\times10^3$fps & 1.833ms & 5.55$\times10^3$fps \\\hline

\end{tabular}
\label{tab:rtx2080-result}
\end{table*}

\begin{table*}[!t]
\centering\scriptsize
\caption{BTC Inference Performance on NVIDIA Turing RTX2080Ti GPU.}
\begin{tabular}{|c|c|c|c|c|c|c|c|c|c|c|c|c|}
\hline
 & \multicolumn{2}{c|}{\bf MNIST-MLP} &  \multicolumn{2}{c|}{\bf Cifar10-VGG} &  \multicolumn{2}{c|}{\bf Cifar10-ResNet14} &  \multicolumn{2}{c|}{\bf ImageNet-AlexNet} &  \multicolumn{2}{c|}{\bf ImageNet-VGG} &  \multicolumn{2}{c|}{\bf ImageNet-ResNet18} \\\hline
 \textbf{Schemes} & \textbf{8 Latency} & \textbf{Throughput} & \textbf{8 Latency}  & \textbf{Throughput} & \textbf{8 Latency} & \textbf{Throughput} & \textbf{8 Latency}  & \textbf{Throughput} & \textbf{8 Latency} & \textbf{Throughput} & \textbf{8 Latency}  & \textbf{Throughput} \\ \hline
\textbf{SBNN-32} & 0.252ms & 2.99$\times10^6$fps & 1.596ms & 1.19$\times10^4$fps & 4.909ms & 4.68$\times10^3$fps & 1.937ms & 7.62$\times10^3$fps & 23.788ms & 5.13$\times10^2$fps & 5.863ms & 3.08$\times10^3$fps \\\hline
\textbf{SBNN-32-Fine} & 0.082ms & 2.43$\times10^6$fps & 1.548ms & 1.19$\times10^4$fps & 4.061ms & 4.65$\times10^3$fps & 1.733ms & 7.16$\times10^3$fps & 22.722ms & 5.06$\times10^2$fps & 5.145ms & 3.10$\times10^3$fps \\\hline
\textbf{SBNN-64} & 0.952ms & 1.03$\times10^5$fps & 2.921ms & 1.37$\times10^4$fps & 8.633ms & 5.95$\times10^3$fps & 14.214ms & 2.59$\times10^3$fps & 31.561ms & 7.08$\times10^2$fps & 8.092ms & 3.91$\times10^3$fps \\\hline
\textbf{SBNN-64-Fine} & 0.070ms & 6.87$\times10^6$fps & 0.926ms & 2.09$\times10^4$fps & 2.341ms & 7.86$\times10^3$fps & 2.017ms & 4.99$\times10^3$fps & 12.057ms & 8.83$\times10^2$fps & 3.233ms & 4.52$\times10^3$fps \\\hline
\textbf{BTC} & 0.057ms & 4.38$\times10^6$fps & 0.317ms & 4.69$\times10^4$fps & 0.728ms & 2.06$\times10^4$fps & 1.878ms & 4.87$\times10^3$fps & 5.840ms & 1.62$\times10^3$fps & 1.538ms & 7.23$\times10^3$fps \\\hline
\textbf{BTC-FMT} & 0.053ms & 6.78$\times10^6$fps & 0.276ms & 5.06$\times10^4$fps & 0.618ms & 2.24$\times10^4$fps & 1.862ms & 4.87$\times10^3$fps & 5.466ms & 1.76$\times10^3$fps & 1.438ms & 7.34$\times10^3$fps \\\hline
\end{tabular}
\label{tab:rtx2080ti-result}
\vspace{0.15truein}
\end{table*} 

Finally, we evaluate the overall BNN implementation. Table~\ref{tab:bhtc_evaluation} lists the 6 models we used for evaluation. Table~\ref{tab:rtx2080-result} and \ref{tab:rtx2080ti-result} list the latency and throughput we obtained for the six models on the two NVIDIA Turing GPUs, respectively. The latency is measured under a batch size of 8 since 8 is the smallest value to leverage the bit-tensorcores so essentially the latency is for the inference of 8 images. The throughput is measured under a batch of 1024 images for \emph{MNIST} and \emph{Cifar10}, and 512 for \emph{ImageNet}. We compare our performance with the four approaches from the latest SBNN work (from \emph{http://github.com/uuudown/SBNN}.) \cite{li2019bstc}. Overall, compared with the best approach \emph{SBNN-64-Fine} from SBNN, our BTC using the default format design achieves  on average 2.10$\times$ in latency and 1.65$\times$ in throughput on RTX2080Ti, and  2.08$\times$ in latency and 1.62$\times$ in throughput on RTX2080 across the six models. Our proposed BTC new format achieves 2.33$\times$ in latency and 1.81$\times$ in throughput on RTX2080Ti, and 2.25$\times$ in latency and 1.77$\times$ in throughput on RTX2080. The best speedup has been achieved by the FSB-format based design on RTX2080Ti GPU for ResNet-14 on Cifar10 --- 3.79$\times$ in latency and 2.84$\times$ in throughput.

Regarding this result, we have three observations: (I) Our BTC design generally achieves more than 2$\times$ over existing work except for \emph{MNIST-MLP} and \emph{ImageNet-Alexnet} where the throughput is actually a little bit worse. The reason is that for \emph{MLP}, a batch of 1024 is still insufficient for fully leveraging the bit-tensorcores, as will be discussed later. For \emph{Alexnet}, the delay of the first layer remains too large (77.4\%) while the other convolution layers are relatively smaller than alternative networks, which cannot fully utilize the BTCs. (II) Although showing better performance, the speedup led by the new FSB format is not as good as in BMM and BConv, the major reason is that both the batch size and the channels are relatively small (batch$\le$1K, channels$\le$512) which is not the region that the FSB format can demonstrate its best speedups (Section~7.2 and 7.3).

\begin{figure*}[!t]
\centering
\vspace{-5pt}
\includegraphics[width=2\columnwidth]{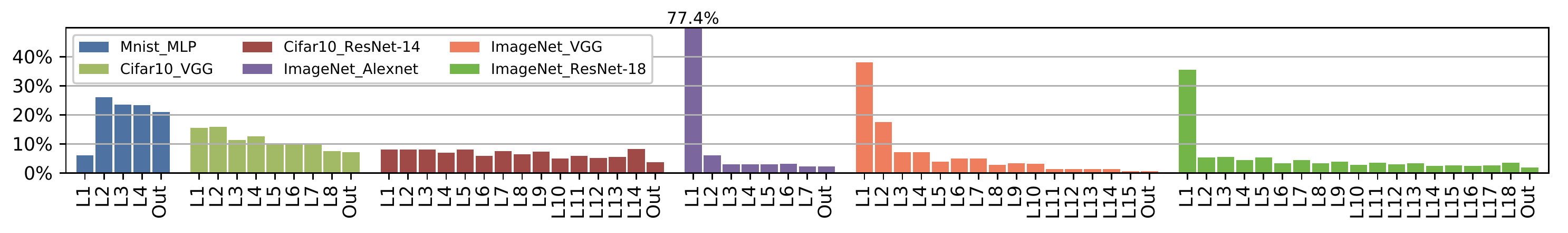} 
\caption{Per-layer latency breakdown of our BTC new-format based BNN design on the 6 models.}
\label{fig:breakdown}
\end{figure*}

\begin{figure*}[!t]
\centering
\includegraphics[width=2\columnwidth]{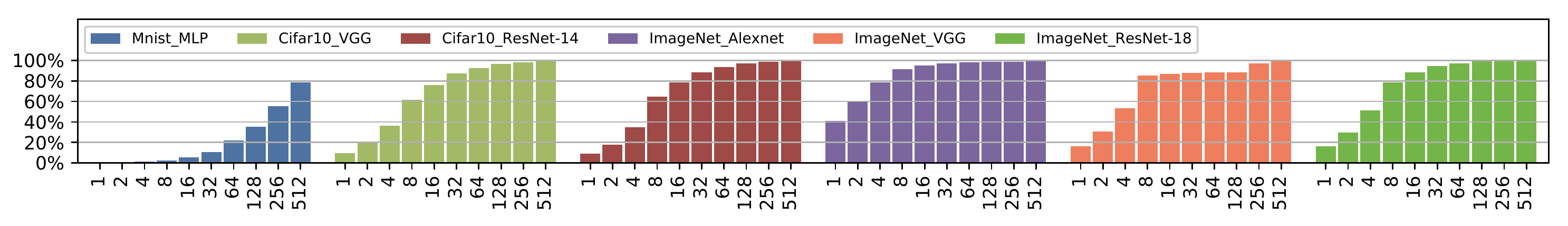} 
\caption{Normalized throughput with respect to increased batch size for the BTC new-format BNN design on the 6 models.}
\label{fig:throughput}
\end{figure*}

\begin{table}[!t]
\centering\scriptsize
\caption{Comparing with FPGA works using AlexNet on ImageNet.}
\begin{tabular}{|c|c|c|c|}
\hline
 \textbf{AlexNet/ImageNet} & \textbf{Platform} & \textbf{Raw Latency} & \textbf{Throughput}   \\ \hline
 RebNet \cite{ghasemzadeh2018rebnet} & Xilinx Virtex VCU108 FPGA &  1902 $\mu s$  & 521 img/s  \\ \hline
 FP-BNN \cite{liang2018fp} & Intel Stratix-V FPGA &  1160 $\mu s$  & 862 img/s  \\ \hline
 O3BNN \cite{geng2019o3bnn} & Xilinx Zynq ZC706 FPGA & 774 $\mu s$ & 1292 img/s  \\ \hline
 SBNN \cite{li2019bstc} & NVIDIA Tesla V100 GPU & 979 $\mu s$ & 4400 img/s  \\ \hline
 BTC & NVIDIA RTX2080Ti GPU &  559 $\mu s$ &  4869 img/s  \\ \hline
 \end{tabular}
\label{tab:compare-alexnet}
\end{table}

\begin{table}[!t]
\centering\scriptsize
\caption{Comparing with CPU, GPU and FPGA using VGG-16 on ImageNet.}
\begin{tabular}{|c|c|c|c|}
\hline
 \textbf{Vgg-16/ImageNet} & \textbf{Platform} & \textbf{Raw Latency} & \textbf{Throughput}   \\ \hline
 BitFlow \cite{hubitflow2018} & NVIDIA GTX1080 &  12.87 ms   & 78  img/s  \\ \hline
 BitFlow \cite{hubitflow2018} & Intel i7-7700 HQ &  16.10 ms   & 62 img/s  \\ \hline
 BitFlow \cite{hubitflow2018} & Intel Xeon-Phi 7210 & 11.82 ms  & 85 img/s  \\ \hline
 O3BNN \cite{geng2019o3bnn}  & Xilinx Zynq ZC706 FPGA & 5.626 ms & 178  img/s  \\ \hline
 SBNN \cite{li2019bstc} & NVIDIA Tesla V100 GPU &  3.208 ms & 312 img/s  \\ \hline
 BTC & NVIDIA RTX2080Ti GPU &  3.570 ms &  1760 img/s  \\ \hline
 \end{tabular}
\label{tab:compare-vgg}
\end{table}

Table~\ref{tab:compare-alexnet} and \ref{tab:compare-vgg} compare the single image raw latency and throughput of our BTC-based new format design with other existing BNN approaches for CPU, GPU, Xeon-Phi and FPGA using Alexnet and VGG-16 on ImageNet. As can be seen, our design achieves the best single-image raw latency and throughput on Alexnet, and more than 5$\times$ throughput enhancement on VGG-16 over the existing works as listed.

\subsection{Sensitivity Study}
We perform several sensitivity studies in this subsection.

\vspace{2pt}\noindent \textbf{Latency Breakdown:} Figure~\ref{fig:breakdown} illustrates the percentage breakdown of the latency (measured by the \emph{clock()} instruction on GPU) for the inference of 8 images over the six models on the RTX-2080 GPU. Clearly, the first layer contributes the most delay for the three ImageNet models due significantly larger image size than the other two datasets. For AlexNet, the percentage can be as high as 77.4\%. It is also over 35\% for VGG-16 and ResNet-18. This is different from existing belief that the first layer is often not a big issue due to the least parameters and computation \cite{hubara2016binarized, li2019bstc}. The latency for other layers are roughly balanced.

\begin{table}[!t]
\centering\scriptsize
\caption{Layer-wise Synchronization Overhead.}
\begin{tabular}{|c|c|c|c|c|c|c|}
\hline
  \textbf{Sync} & {Mnist-MLP} & {Cifar10-vgg} & {Cifar10-ResNet14} &  {Alexnet} & {VGG} & {ResNet18} \\\hline
  \% & 8.26\% &  14.10\% & 13.20\% & 1.36\% & 1.72\% & 5.67\%  \\ \hline
 \end{tabular}
\label{tab:sync}
\end{table} 

\vspace{2pt}\noindent \textbf{Synchronization Overhead:} As we enforce global synchronization through cooperative-groups per layer to ensure data consistency, such global synchronizations can introduce extra overhead and idle waiting of SMs. Table~\ref{tab:sync} shows the percentage of this synchronization overhead, which is measured by removing all the synchronization primitives. As can be seen, this overhead is the most for the medium network models, e.g., the two on Cifar10 (14.1\% and 13.2\%).

\begin{figure}[!t]
\centering
\includegraphics[width=0.95\columnwidth]{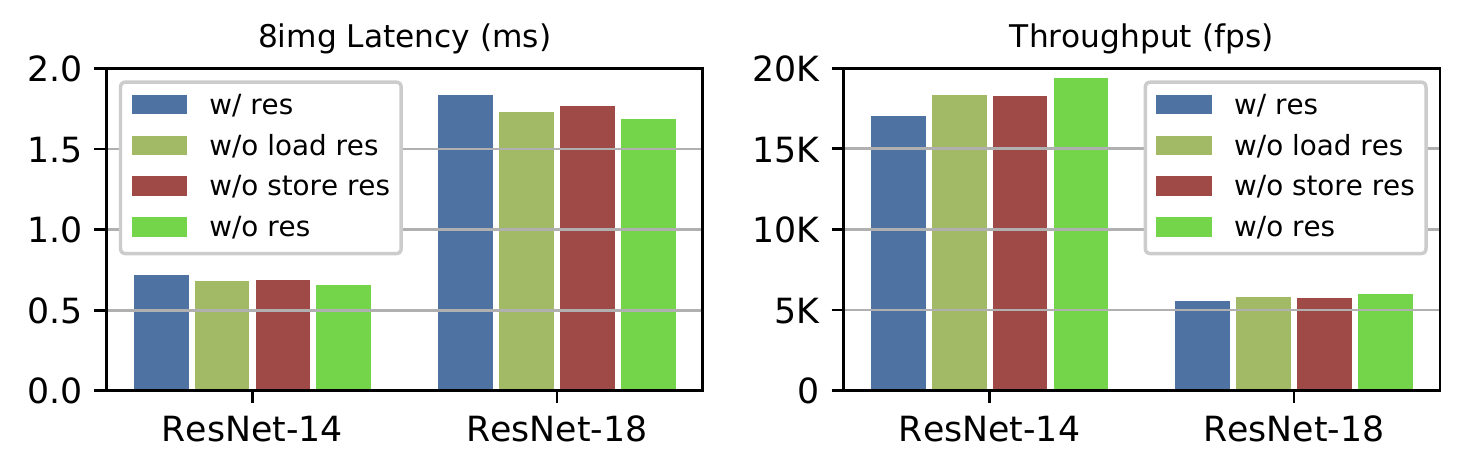} 
\caption{Cifar-ResNet14 Latency: 9.7\% speedup, Throughput: 14\%, ImageNet-ResNet18 Latency: 9.0\% speedup, Throughput: 8.3\% }
\label{fig:residual}
\end{figure}

\vspace{2pt}\noindent \textbf{Shortcut Overhead:} We then focus on the two ResNet models and measure the overhead incurred by handling the cross-layer residual. Figure~\ref{fig:residual} show the latency and throughput of the two ResNet models on RTX-2080 regarding four scenarios: (a) with residual; (b) save the residual without fetching them; (c) fetch the residual without saving them; and (d) without the residual at all. 
For ResNet-14 on Cifar10, if eliminating the residual-related operations, we can gain 9.7\% speedup in latency and 14\% in throughput. For ResNet-18 on ImageNet, we gain 9.0\% and 8.3\%.

\vspace{2pt}\noindent \textbf{Utilization:} We investigate the impact of batch size over the throughput. If the batch size is too small, the hardware such as the bit-tensorcores might be under-utilized. Figure~\ref{fig:throughput} shows the inference throughput of the six models with different batch sizes (normalized to the throughput with a batch of 1024 for MNIST and Cifar10, and 512 for ImageNet) on RTX2080. As can be seen, for ImageNet, a batch of 128 is sufficient to achieve the best throughput while for Cifar10, a batch of 512 is necessary. For MNIST, even with batch size arising from 16K to 32K, the throughput is still increasing. The best throughput is obtained at 32K with 7.62$\times10^6$ fps.

\begin{table}[!t]
\centering\scriptsize
\caption{\blue{8img Latency for ResNet with increased depth on RTX2080.}}
\begin{tabular}{|c|c|c|c|c|}
\hline
  {ResNet/ImageNet} & {ResNet-18} & {ResNet-50} & {ResNet-101} & {ResNet-152} \\\hline
  BTC & 1.538 ms & 4.668 ms & 8.873 ms & 13.393 ms   \\ \hline
  BTC-FMT & 1.438 ms & 4.169 ms & 8.521 ms & 13.316 ms   \\ \hline
 \end{tabular}
\label{tab:deep}
\end{table} 

\vspace{2pt}\noindent \blue{\textbf{Network Depth:} Finally we look at the performance impact from increased network depth. Table~\ref{tab:deep} lists the raw latency for 8 image inference with ResNet-18, 50, 101 and 152 \cite{he2016deep} on RTX2080. As can be seen, the latency increases almost in linear with more layers in the network. Even with very deep models, our BNN framework can still work well.}

\subsection{\blue{BTC-based BENN Scaling}}

\begin{figure}[!t]
\centering
\includegraphics[width=0.9\columnwidth]{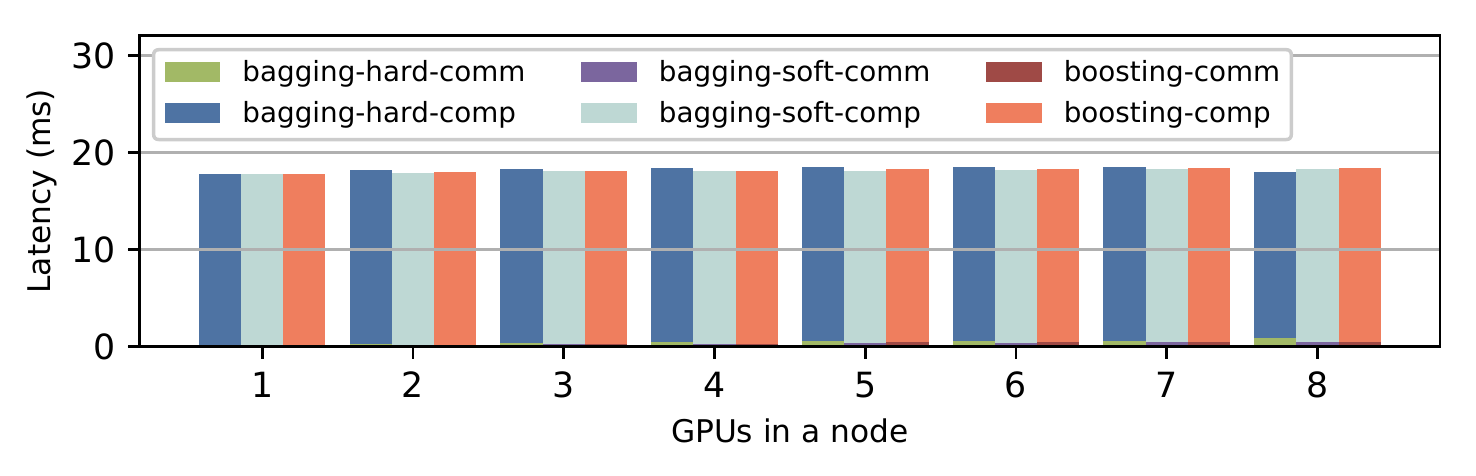} 
\caption{\blue{Latency breakdown for BENN scaling-up with more GPUs in a node.}}
\label{fig:benn_scaleup}
\end{figure}

\begin{figure}[!t]
\centering
\includegraphics[width=0.9\columnwidth]{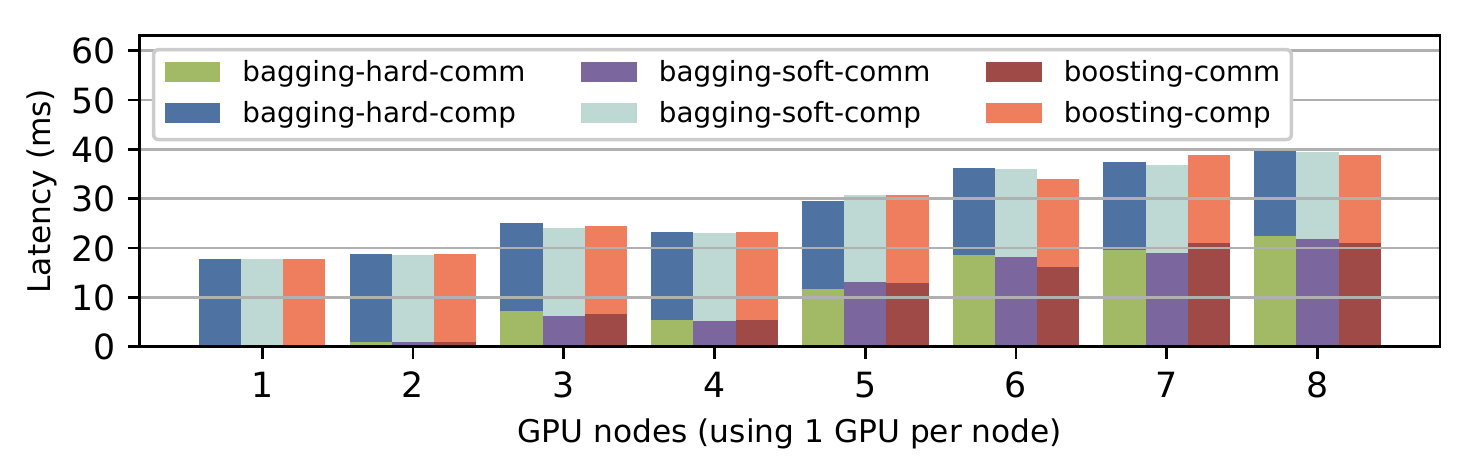} 
\caption{\blue{Latency breakdown for BENN scaling-out with more GPU nodes.}}
\label{fig:benn_scaleout}
\end{figure}

\blue{To compensate potential accuracy loss of BNNs, Zhu et al. recently proposed BENNs \cite{zhu2018binary}, which assembled multiple BNNs through bagging and boosting to tackle the intrinsic instability of BNN training, thus achieving superior accuracy, e.g., top-1 54.3\% for AlexNet and 61\% for ResNet-18 on ImageNet). We implement the BTC-based BENN which executes each BENN's BNN component on a single GPU concurrently and merges their output through collective communication over the inter-node network or intra-node GPU interconnects \cite{li2019evaluating}. We evaluate our design on an HPC cluster with 8 nodes connected by InfiniBand (IB), and each node incorporates 8 NVIDIA RTX-2080Ti GPUs (see Table~\ref{tab:gpu}) connected by \emph{PCI-e} \cite{li2018tartan}.}

\blue{We perform two types of evaluations: (i) \emph{Scaling-up:} we use a single node, and increase the number of BNN components inside BENN, corresponding to adopting more GPUs in the execution. We use the \emph{reduction} collective communication primitive from the NCCL library \cite{nccl} for the bagging and boosting operations of BENN. Figure~\ref{fig:benn_scaleup} illustrates the latency breakdown of BENN inference using the \emph{hard-bagging}, \emph{soft-bagging} and \emph{boosting} ensemble methodologies \cite{zhu2018binary}. Each BENN component is a BTC-based ResNet-18 BNN using the new FSB format. The batch size is 128. As can be seen, due to the small amount of data transfer and the high efficiency of NCCL, the communication overhead is tiny for all the three ensemble techniques, which implies that by adopting BENN, we can greatly improve BNN accuracy but still gain the great performance advantage of BNN. (ii) \emph{Scale-out:} in this test, we adopt all the 8 nodes, but use only a single GPU per node for the BENN inference. We use \emph{MPI\_reduce} primitive from \emph{Intel-MPI-2019u4} for the bagging and boosting operations. As shown in Figure~\ref{fig:benn_scaleup}, different from the scaling-up, with increased BENN components, the latency surges when scaling-out. With 8 GPUs, the communication latency is even higher than the BNN inference itself. This suggests that communication is key to BENN design.}

%\subsection{\blue{Evaluation Summary}}
%\blue{To summarize, the evaluation results on the two NVIDIA Turing GPUs show that: (a) The BTC-based BMM design can achieve up to 20$\times$ over the 32-bit full-precision cuBLAS GEMM, and up to 2.5$\times$ over state-of-the-art BMM approach on RTX2080; (b) The BTC-based BConv design can achieve up to 25$\times$ over the 32-bit full-precision cuDNN Conv, and up to 3.1$\times$ over state-of-the-art BConv approach on RTX2080Ti; (c) For BNN inference, our BTC-based new design can achieve generally 2$\times$ compared with the best reference design. Regarding the execution, the full-precision first layer, the per-layer synchronization, and the shortcut structure account for up to 77\%, 14\% and 12\% overhead.}

% (d) Boosting via multi-GPUs can achieve better accuracy with negligible overhead.

\subsection{Discussion on BNN Accuracy and Usage}
\blue{We further discuss the accuracy of BNN compared to DNN. So far, all the concerns regarding the accuracy loss of BNNs are based on two implicit assumptions: (a) BNN and DNN are using the same number of neurons and channels; (b) BNN and DNN are using the same models that were originally designed for full-precision DNN. For (a), if the number of neurons or channels for BNN can be expanded, e.g., through boosting, we can gain compatible or even better accuracy than DNN \cite{zhu2018binary} with little overhead, as already shown. For (b), the network structure designed for DNN (e.g., VGG block) might not be the most suitable for BNNs. Recently, people started to look at BNN-oriented network structure, such as the \emph{dense block} + \emph{improvement block} structure in MeliusNet \cite{bethge2020meliusnet}. With such novel structures, BNN's accuracy can be greatly improved. We anticipate BNN to be quite useful for (i) (Embedded) real-time system (e.g., \cite{chen2020gpu}); (ii) Large-scale data preprocessing and vision (e.g., \cite{say2018planning, korneev2018constrained, ma2018binary}). Finally, BNN is continuously a focused area in the ML community. More advanced BNN-specific models and training methods are under active development.}

%\subsection{BNN Accuracy}
%Why BNN can work? Why BNN accuracy? Why MeliusNet can work?
%\subsection{BNN for HPC}
%What BNN can be used for in HPC?
%Parsing indexed data, for example, in data triggering, HEP data fetching, document embedding.
%\subsection{BTC for other Utilization}
%Potentially for other approaches such as Graph BLAST.

\section{Conclusion}
In this paper we investigate the new bit computation capability of the tensorcores in Turing GPUs. We found that the stride of memory access can significantly impact the performance of memory access. Based on this observation, we propose a new bit data format for efficient design of Bit-Matrix-Multiplication and Bit-Convolution. We built the full implementation for BNN inference. Evaluations using six network models (MLP, VGG-like, AlexNet, VGG-16, ResNet-14/18) on three datasets (MNIST, Cifar10 and ImageNet) over two latest Turing GPUs (RTX2080 and RTX2080Ti) show that our design can bring on average 2.33$\times$ (up to 3.79$\times$) in latency and 1.81$\times$ (up to 2.84$\times$) in throughput compared with state-of-the-art BNN design for GPUs, leading to super realtime performance. Future work include: (a) Exploiting BTC for alternative utilization such as Graph-BLAS computation; (b) Investigating a combination of FPGAs/ASICs and GPUs for BNN online training tasks; (c) Testing in Ampere GPUs.

\section*{Acknowledgments}
We thank the anonymous reviewers for their insightful feedback. This research was supported by PNNL's DMC-CFA  and DS-HPC LDRD projects. This research was supported by the U.S. DOE SC, ASCR, under award 66150: "CENATE - Center for Advanced Architecture Evaluation".

%PNNL is operated by Battelle for the U.S. Department of Energy (DOE) under contract DE-AC05-76RL01830.

%The Pacific Northwest National Laboratory is operated by Battelle for the U.S. Department of Energy (DOE) under contract DE-AC05-76RL01830.

%\ifCLASSOPTIONcompsoc
%\else
%\fi


\bibliographystyle{unsrt}
%\bibliographystyle{IEEEtranS}
%\bibliography{references}

{\scriptsize
\bibliography{bare_jrnl_compsoc}}

\ifCLASSOPTIONcaptionsoff
  \newpage
\fi

\end{document}